\documentclass{article}
\usepackage[utf8]{inputenc}
\usepackage{float}
\usepackage{jheppub}
\usepackage{amsmath}
\usepackage[english]{babel}
\usepackage{amssymb}
\usepackage{mathtools}
\usepackage{color}
\definecolor{ao}{rgb}{0.0, 0.5, 0.0}

\usepackage{graphicx}
\usepackage{longtable}
\usepackage{tikz}
\usepackage{comment}
\usepackage{capt-of}
\usepackage{amsthm}
\usepackage{hyperref}
\hypersetup{
    colorlinks=true,
    linkcolor=blue
    }

\usepackage{bm}

\usepackage{caption}
\usepackage{subcaption}

\newcommand{\old}[1]{\textcolor{black}{#1}}
\newcommand{\new}[1]{\textcolor{black}{#1}}

\newtheorem{remark}{Remark}

\newtheorem{proposition}{Proposition}
\newtheorem{corollary}{Corollary}
\newtheorem{definition}{Definition}
\newtheorem{lemma}{Lemma}
\newtheorem{assumption}{Assumption}

\def\0{\mbox{\tiny $0$}}
\def\1{\mbox{\tiny $1$}}
\def\2{\mbox{\tiny $2$}}
\def\3{\mbox{\tiny $3$}}
\def\4{\mbox{\tiny $4$}}
\def\5{\mbox{\tiny $5$}}
\def\6{\mbox{\tiny $6$}}
\def\7{\mbox{\tiny $7$}}
\def\8{\mbox{\tiny $8$}}
\def\9{\mbox{\tiny $9$}}
\def\M{\mbox{\tiny $M$}}
\def\chiM{\hat{\chi}_{\M}}
\def\etaM{\hat{\eta}_{\M}}

\def\k{k_{_{B}}}

\def\r{\rangle}
\def\R{\mathcal{R}}
\def\l{\langle}
\def\m{\bar{m}}
\def\V{\mathcal{V}}

\def\q{\bar{q}}
\def\n{\bar{n}}
\def\nRS{\bar{n}'}

\def\qqq{\tilde q_2}
\def\b{\beta^{'}}

\newcommand{\SOMMA}[2]{\displaystyle\sum\limits_{#1}^{#2}}
\newcommand{\sommaSigma}[1]{\displaystyle\sum\limits_{\lbrace#1\rbrace}}

\newcommand{\D}{{\mathcal{D}}}

\long\def \beq#1\eeq {\begin{equation} #1 \end{equation}}
\long\def \beaq#1\eeaq {\begin{equation}\begin{aligned} #1 \end{aligned}\end{equation}}
\long\def \bes#1\ees {\begin{equation}\begin{split} #1 \end{split} \end{equation}}
\long\def \bea#1\eea {\begin{eqnarray} #1 \end{eqnarray}}
\long\def \bse[#1]#2\ese {\begin{subequations}\label{#1}\begin{align} #2 \end{align}\end{subequations}}

\newcommand{\sums}{\sum_{ \boldsymbol \sigma}}

\newcommand{\si}{\sigma_i}

\newcommand{\bQ}{{\bm Q}}

\newcommand{\hence}{\Rightarrow}

\title{Unsupervised and Supervised learning by Dense Associative Memory under replica symmetry breaking}
\author[1,2,3,5]{Linda Albanese,}
\author[4,5]{Andrea Alessandrelli,}
\author[2]{Alessia Annibale,}
\author[1,5]{Adriano Barra}

\affiliation[1]{Dipartimento di Matematica e Fisica ``Ennio De Giorgi'', Universit\`a del Salento, Lecce, Italy.}
\affiliation[2]{Department of Mathematics, King's College London, The Strand, London, UK.}
\affiliation[3]{Scuola Superiore ISUFI, Universit\`a del Salento, Lecce, Italy.}
\affiliation[4]{Dipartimento di Informatica, Universit\`a di Pisa, Pisa Italy.}
\affiliation[5]{Istituto Nazionale di Fisica Nucleare, Sezione di Lecce, Italy.}

\abstract{Statistical mechanics of spin glasses is one of the main strands toward a comprehension of information processing by neural networks and learning machines. Tackling this approach, at the fairly standard replica symmetric level of description, recently Hebbian attractor networks with multi-node interactions (often called Dense Associative Memories) have been shown to outperform their classical pairwise counterparts in a number of tasks, from their robustness against adversarial attacks and their capability to work with prohibitively weak signals  to their supra-linear storage capacities. 
Focusing on mathematical techniques more than computational aspects, in this paper we relax the replica symmetric assumption and we derive the one-step broken-replica-symmetry picture of supervised and unsupervised learning protocols for these Dense Associative Memories: a phase diagram in the space of the control parameters is achieved, independently, both via the Parisi's hierarchy within then replica trick as well as via the Guerra's telescope within the broken-replica interpolation.  Further, an explicit analytical investigation is provided to deepen both the {\em big-data} and {\em ground state} limits of these networks as well as a proof that replica symmetry breaking does not alter the thresholds for learning and slightly increases the maximal storage capacity.  Finally the De Almeida and Thouless line, depicting the onset of instability of a replica symmetric description, is also analytically derived highlighting how, crossed this boundary, the broken replica description should be preferred.}

\abstract{Hebbian neural networks with multi-node interactions, often called Dense Associative Memories, have recently attracted considerable interest in the statistical mechanics community, as they have been shown to outperform their pairwise counterparts in a number of features, including resilience against adversarial attacks, pattern retrieval with extremely weak signals and supra-linear storage capacities. However, their analysis has so far been carried out within a replica-symmetric theory. In this manuscript, we relax the assumption of replica symmetry and analyse these systems at one step of replica-symmetry breaking, focusing on two different prescriptions for the patterns that we will refer to as supervised and unsupervised learning. We derive the phase diagram of the model using two different approaches, namely Parisi's hierarchical ansatz for the relationship between different replicas within the replica approach, and the so-called telescope ansatz within Guerra's interpolation method: our results show that replica-symmetry breaking does not alter the threshold for learning and slightly increases the maximal storage capacity. Further, we also derive analytically the instability line of the replica-symmetric theory, using a generalization of the De Almeida and Thouless approach.}

\begin{document}

\maketitle

\section{Introduction}


Since Hopfield's seminal work on the use of 
biologically inspired neural networks 
for associative memory and pattern recognition tasks \cite{Hopfield}, there have been many important contributions applying concepts from statistical mechanics, in particular spin-glass theory, to the study of neural networks with pairwise Hebbian interactions (for an overview of the vast literature, the reader is referred to the books \cite{MPV, TalaBook, Amit, Dotsenko, Engel,  Nishimori, CoolenKuhnSollich,
Huang}
and to the recent reviews \cite{Lenka,Carleo}). This  connection was first pointed out by Hopfield himself in \cite{Hopfield}, where he showed that neural networks with pairwise Hebbian interactions were particular realizations of spin glasses, and it was further developed by Amit, Gutfreund, and Sompolinsky \cite{AGS} who showed that the free-energy landscape of such systems is characterized by a large number of local minima, 
corresponding to different patterns of information, retrieved by the network
from different initial conditions.
%
In addition to neural networks with pairwise Hebbian interactions, a large body of work 
has focused on 
extending Hebbian learning to multi-node interactions, both in the early days (see e.g. \cite{kanter1988asymmetric, Gardner, Baldi}) and more recently (see e.g. \cite{HopKro1, ramsauer2020hopfield}). These models are often referred to as \textit{dense} associative memories \cite{HopKro1}, as they can store many more patterns than the number of neurons in the network. In particular, they can perform pattern recognition with a supra-linear storage load \cite{Baldi} and can work at very low signal-to-noise ratio, when compared to their pairwise counterparts \cite{Barra-PRLdetective}. In addition, they have been shown to be resilient to adversarial attacks \cite{Krotov2018} and 
\old{combinations of such models can lead to exponential storage capacity \cite{Lucibello2023}}.  

In recent years, a duality between Hebbian neural networks
and machine learning models has been pointed out. For example, the Hopfield
model has been shown to be equivalent to a restricted Boltzmann machine \cite{Contucci}, an archetypical model for
machine learning \cite{hinton1984boltzmann}, and sparse restricted Boltzmann machines have been mapped to Hopfield models with diluted patterns \cite{parallel1, parallel2}.
Furthermore, restricted Boltzmann machines with generic priors 
have led to the definition of generalized Hopfield models \cite{Remi, DaniPRE2017, DaniPRE2018} and
neural networks with multi-node Hebbian interactions have recently been shown to be equivalent to higher-order Boltzmann machines \cite{AAAF-JPA2021, agliariAlone} and deep Boltzmann machines 
\cite{alberici2020annealing, alberici2021deep}. As a result, multi-node Hebbian learning is receiving a second wave of interest since its foundation in the eighties \cite{kanter1988asymmetric, Gardner} 
as a paradigm to understand  deep learning \cite{ramsauer2020hopfield, Metha2014}.

In order to make this connection 
clearer, we should stress that neither
the Hopfield model nor its generalized versions with multi-node Hebbian interactions, are machine learning models, in that they are not trained from data and their couplings do not evolve in time according to a learning rule. Instead, their couplings are fixed from the outset to the values that {\it would have been attained} by a neural network trained according to the Hebbian rule, which adjusts the strengths of the connections between the neurons based on the input patterns that the network is exposed to.
In this work, we consider two different routes for exposing the network to 
data, as considered in \cite{EmergencySN, prlmiriam}. In particular, we analyze the scenario where the  
input patterns are noisy or corrupted versions of pattern archetypes.  
In the first route, that we will call `unsupervised', we expose the network to all of the input patterns together, without specifying the archetype that each pattern is meant to represent \cite{barlow1989unsupervised}.  In the second route, that we will call `supervised', we split the training dataset in different classes, corresponding to the archetypes, and the network is exposed to the examples in each class, class by class, in the same way that a teacher helps a student to learn one topic at time, from different examples \cite{cunningham2008supervised}. 
Beyond being more realistic than traditional Hebbian storing, this generalization allows to pose a number of new questions, such as: 
is the network able, in either scenario, to retrieve the archetypes by itself, i.e. find hidden and intrinsic structures in the training dataset?
which is the minimal amount of examples per archetype that the network has to experience, given the noise in the examples and the number of archetypes, in the two scenarios?

These questions have been addressed in neural networks with pairwise as well as multinode Hebbian interactions, for both random uncorrelated datasets and structured datasets (including the well-known MNist or Fashion MNist) \cite{prlmiriam, unsup, super},
within a replica-symmetric (RS) analysis, which is based on the assumption that different replicas (or copies) of the systems 
are invariant under permutations. However, since neural networks 
are particular realization of spin-glasses, such symmetry is expected to be broken in certain regions of the parameters space, where the system develops a multitude of degenerate states. In a recent work \cite{albanese2023almeida}, we devised 
a simple and rigorous method to detect the onset of the instability of RS theories in neural networks with pairwise and multinode Hebbian interactions. Building on this work, we derive the RS instability lines for the two different protocols defined here, supervised and unsupervised, and we re-investigate the problem defined above, previously analysed within an RS assumption, by carrying out the analysis at one step of replica-symmetry breaking ($1$RSB). 
Since Parisi's seminal work on replica symmetry breaking \cite{MPV}, several alternative mathematical techniques have been proposed to investigate replica-symmetry broken phases \cite{bovier2012mathematical, gradenigo2020solving, barra2010replica}. In this work, 
we will use 
interpolations techniques, that were pioneered by Guerra in the context of his work on spin glasses \cite{guerra_broken} and were later applied to neural networks (e.g. \cite{AABO-JPA2020}). 
One advantage of this approach is that it avoids certain heuristics required by the replica method and it leads to simpler calculations. 
For completeness, we will also derive the same results using the replica approach, with the pedagogical aim of creating a bridge between the two methods,  
one being a cornerstone of the statistical physics community (Parisi's), the other constituting a golden niche in the  
mathematical physics community (Guerra's).
\par\medskip
The paper is structured as follows.
In Section \ref{sec:unsup} we define a first model for dense associative memories that we will shall refer to as 'unsupervised' and we analyse it at one step of replica symmetry breaking via Guerra's interpolation method and replica techniques.
 In Section \ref{sec:sup} we define a second model for dense associative memories, that we will shall refer to as 'supervised' and we carry out the same analysis as in Sec. \ref{sec:unsup}. 
Finally, in Sec. \ref{sec:conclusion} we summarize and discuss our results. We relegate technical details to the Appendices as well as the derivation of the instability line of the RS theory, which shows the importance of analysing these systems under the assumption of replica symmetry breaking.

\section{`Unsupervised' dense associative memories}
\label{sec:unsup}

In this section we analyse the information processing capabilities of Dense Associative Memories (DAM) in the so-called `unsupervised' setting, as introduced in Sec. \ref{sec:unsup_gen}, via 
two different approaches, namely Guerra's interpolation techniques (Sec. \ref{GuerraUnsup}) and Parisi's replica approach (Sec. \ref{ParisiUnsup}). Both analyses are carried out at the first step of replica-symmetry breaking. Results are discussed in Sec. \ref{sec:results-unsup}. 
The instability line of the RS theory (equivalent to the de Almeida and Thouless line in the context of spin-glasses), showing the importance of working under the assumption of broken replica symmetry, is derived in Appendix \ref{sec:AT_unsup}.
 
\subsection{Model and definitions}
\label{sec:unsup_gen}

We consider a system of $N$ neurons, modelled via Ising spins $\si \in  \{ -1, 1 \}$, with  $i= {1,\hdots ,N}$, 
where neurons interact via $P$-node interactions. 
We assume to have $K$ Rademacher archetypes, 
defined as $N$-dimensional vectors 
$\bm \xi^{\mu}$, with $\mu =1,\hdots, K$, where each entry  is drawn randomly and independently from the distribution 
\begin{align}
\label{eq:xi}
    \mathbb{P}(\xi_i^\mu) = \dfrac{1}{2} \left( \delta_{\xi_i^\mu,+1} + \delta_{\xi_i^\mu,-1} \right).
\end{align}
Such archetypes are not provided to the network, 
instead the network will have to infer them by experiencing solely noisy or corrupted versions of them.
In particular, we assume that for each archetype $\mu$, $M$ examples $\bm\eta^{\mu,a}$, with $a=1, \hdots, M$, are available, which are corrupted versions of the archetype, such that
\begin{equation} \label{eq:Bernoulli}
    \mathbb{P}(\eta_{ i}^{\mu,a}|\xi_i^\mu) = \frac{1-r}{2} \delta_{\eta_{ i}^{\mu,a},-\xi_i^\mu} + \frac{1+r}{2} \delta_{\chi_{ i}^{\mu,a},+\xi_i^\mu},
\end{equation}
where $r \in [0,1]$ controls the \emph{quality} of the dataset, i.e. for $r =1$ the example matches perfectly the archetype, while for $r=0$ it is purely stochastic. To quantify the information content of the dataset it is useful to introduce the variable 
\begin{equation}
\rho =\frac{1-r^2}{Mr^2} 
\label{eq:rho}
\end{equation}
that we shall refer to as the \textit{dataset entropy}.\footnote{It was 
shown in \cite{prlmiriam} that the conditional entropy which quantifies the amount of information needed to reconstruct the original pattern $\bm\xi^\mu$ given the set of related examples $\{\eta^{\mu,a}\}_{a=1,\hdots,M}$, is a monotonically increasing function of $\rho$.}
We note that $\rho$ vanishes either 
when the examples are identical to the archetypes (i.e. $r \to 1$) or when the number of examples is infinite (i.e. $M \to \infty$), or both. 
If we regard the traditional DAM model as storing patterns and the unsupervised DAM model as learning patterns from stored examples, then 
$\rho$ can be thought of as the parameter that 
quantifies the difference between storing and 
learning (which is larger, the larger $\rho$).


\begin{definition} \label{def:dense_unsup}
The cost function (or \textit{Hamiltonian}) of the `unsupervised' DAM model is 
\begin{align}
    \mathcal{H}_{N}^{(P)}(\bm \sigma \vert \bm \eta) =& -\dfrac{1}{\,\R^{P/2}\,M\,N^{P-1}}\SOMMA{\mu=1}{K}\SOMMA{a=1}{M}\left(\SOMMA{i_1 < \cdots < i_P}{N,\cdots, N}\eta_{i_1}^{\mu,a }\cdots\eta_{i_P}^{\mu,a }\sigma_{_{i_1}}\cdots\sigma_{_{i_P}}\right),
    \label{def:H_PHopEx}
\end{align}
where the constant $\R^{P/2}$, with $\R=r^2(1+\rho)$ and $\rho$ defined in \eqref{eq:rho},
is included 
in the definition 
for mathematical convenience, while the factor $N^{P-1}$ ensures that the Hamiltonian is extensive in the network size $N$. 
On the left hand side (LHS), the label $(P)$ denotes the order of the multi-node interactions and $\bm\eta$ is a short-hand for the collection of 
all the examples, for all the archetypes 
$\{\bm\eta^{\mu,a}\}_{\mu=1,a=1}^{K,M}$. 

\end{definition}

\begin{remark}
    For $P=2$, i.e. pairwise interactions, the 
    unsupervised DAM model reduces to the Hopfield neural network in the unsupervised setting \cite{EmergencySN}. In addition, in the absence of dataset corruption, i.e. $r=1$, $\mathcal{R} =1$ and the model reduces to the standard Hopfield model, as analysed in  \cite{AGS, Amit}.
\end{remark}
\begin{definition}
The partition function associated to the Hamiltonian \eqref{def:H_PHopEx} at inverse noise level $\beta\in \mathcal{R}^+$
is defined as 
\begin{align}  \mathcal{Z}^{(P)}_{N}(  \bm \eta) = \sum_{\bm \sigma}^{2^N} \exp \left( -\beta \mathcal{H}^{(P)}_{N}(\bm \sigma \vert \bm \eta)\right)=: \sum_{\bm \sigma}^{2^N} \mathcal{B}^{(P)}_{N}(\bm \sigma  \vert \bm \eta),
    \label{eq:Def_orig_Z}
\end{align}
where $\mathcal{B}^{(P)}_{N}(\bm \sigma \vert \bm \eta)$ is referred to as the Boltzmann factor. 
At finite network size $N$, the free energy of the model $\mathcal{F}^{(P)}_{N}$ is given by 
    \begin{align}
        \mathcal -\beta \mathcal{F}^{(P)}_{N} = \frac{1}{N}\mathbb{E}\log \mathcal{Z}^{(P)}_{N}(\bm \eta)
        \label{PressureDef_unsup}
    \end{align}
    where $\mathbb{E}[.]$ denotes the average over the realizations of examples $\bm\eta$, regarded as quenched.
 \end{definition}
We are interested in the behaviour of the system in the limit of large system size $N\to \infty$ and finite network load $\alpha$, as specified by the following
\begin{definition}
    In the thermodynamic limit $N \to \infty$, the load of the unsupervised DAM model is defined as
\begin{equation}
\lim_{N\to +\infty} \frac{K}{N^{P-1}} =: \alpha < \infty
\label{eq:carico_TDL_unsup}
\end{equation}
We will mostly focus on the so-called `saturated' regime, where $\alpha>0$. The regime where the system is away from saturation can be inspected by taking the limit $\alpha\to 0$.
For convenience, we also introduce the parameter $\gamma$, defined by
\begin{equation}
    \alpha=\gamma \dfrac{2}{P!}.
    \label{eq:alphaPP-1}
\end{equation}

The free energy in the thermodynamic limit is denoted as
\begin{equation} \label{eq:statpress_LTD}
\mathcal F^{(P)}  = \lim_{N \to \infty} \mathcal F^{(P)}_{N} .
\end{equation}
\end{definition}

In the following, we focus on the ability of the network to retrieve a single archetype, 
say $\nu$. Given the invariance of the Hamiltonian w.r.t. permutations of the archetypes, we will set without loss of generality $\nu=1$.
Then, following standard procedures \cite{Amit}, we split the sum over $\mu$ in the Hamiltonian into the contribution from $\mu=1$ (regarded as the signal) and the contributions from the other archetypes $\mu>1$ (regarded as slow noise).
Furthermore, we add in the argument of the exponent of the Boltzmann factor a term $J \sum_i \xi_i^1 \si$, so that the free energy can 
serve as a moment generating functional of the so-called 
Mattis magnetization $m_1$ (\textit{vide infra}) by taking derivatives of $\ln\mathcal Z_N^{(P)}$ w.r.t. $J$. As this term is not part of the original Hamiltonian, $J$ will be set to zero later on. 
\newline
Starting from \eqref{eq:Def_orig_Z}, the resulting partition function is
\begin{align} \label{eq:integral}
    \mathcal{Z}^{(P)}_{N} (\bm \eta) =&  \lim_{J \rightarrow 0} \sum_{\bm \sigma}  \exp \left[ J \sum_{i=1}^N \xi_i^1 \si + \dfrac{\beta '}{2\,\R^{P/2} M\,N^{P-1}}\SOMMA{a=1}{M}\left(\sum_{i=1}^N\eta_{_{i}}^{a,1}\sigma_{_{i}}\right)^{^P} \right. \notag 
    \\
    &\left.+\dfrac{\beta 'P!}{2\R^{P/2} M\,N^{P-1}} \, \SOMMA{\mu>1}{K}\SOMMA{a=1}{M} \SOMMA{ \ {i_1 < \cdots < i_P}}{N,\cdots,N}\left(\eta^{\mu\,,a}_{i_1}\cdots\eta^{\mu\,,a}_{i_{P}}\:\sigma_{_{i_1}}\cdots\sigma_{_{i_{_{P}}}}\right)\,\right]
\end{align}

where we have set $\beta'= 2\beta/ P!$ and for the first pattern ($\mu=1$) we used
\new{$P!\sum_{i_1 <\cdots < i_P}=\sum_{i_1\neq \cdots\neq i_P}$
and neglected terms which vanish in the thermodynamic limit}. 
Focusing only on the last term in \eqref{eq:integral} 
we note that the product $\eta_{i_1}^{\mu,a}\ldots \eta_{i_P}^{\mu,a}$ is a random i.i.d. Binomial variable for each $\mu, a$, and the sum over $a$ converges, by the central limit theorem (CLT), to a Gaussian variable with suitably defined mean and variance, i.e. 
\begin{equation}
        \left(\dfrac{1}{M}\SOMMA{a=1}{M}\eta^{\mu\,,a}_{i_1}\cdots\eta^{\mu\,,a}_{i_{P}}\right)\sim \mu+\lambda^\mu_{i_1,\hdots,i_P} \frac{\sigma}{\sqrt{M}},\quad\quad\quad \lambda_{i_1\ldots i_P}^\mu\sim {\mathcal N}(0,1),
\end{equation}
where
\bea
\mu=\langle \eta^{\mu}_{i_1}\cdots\eta^{\mu}_{i_{P}}\rangle_{\bm{\xi}}
= \left(r\xi_{i_1}^\mu\right)\ldots \left(r\xi_{i_P}^\mu\right)
\eea
and 
\bea
\sigma=\sqrt{\langle (\eta^{\mu}_{i_1}\cdots\eta^{\mu}_{i_{P}})^2\rangle_{\bm{\xi}}-\langle \eta^{\mu}_{i_1}\cdots\eta^{\mu}_{i_{P}}\rangle_{\bm{\xi}}^2}
=\sqrt{(\xi_{i_1}^\mu)^2\ldots (\xi_{i_P}^\mu)^2
    -r^{2P}(\xi_{i_1}^\mu\ldots \xi_{i_P}^\mu)^2
    }
\eea
due to the independence of different sites. 
This enables us to write
\begin{eqnarray}
    \label{eq:con_V}
    \left(\dfrac{1}{M}\SOMMA{a=1}{M}\eta^{\mu\,,a}_{i_1}\cdots\eta^{\mu\,,a}_{i_{P}}\right)&\sim& 
    r^P\left(1+\lambda_{i_1\ldots i_P}^\mu\sqrt{\rho_P}
    \right)\xi_{i_1}^\mu\ldots \xi_{i_P}^\mu
\end{eqnarray}
with
$$
\rho_P=\frac{1-r^{2P}}{Mr^{2P}}.
$$
Similarly, applying the CLT to the sum over $\mu$ we get
$$
\left(\dfrac{1}{M N^{P-1}}\sum_a \sum_\mu \eta^{\mu\,,a}_{i_1}\cdots\eta^{\mu\,,a}_{i_{P}}\right)
= \alpha\sqrt{\frac{r^{2P}}{K}}(1+\sqrt{\rho_P})\lambda_{i_1\ldots i_P},
$$
having used 
$$
\frac{1}{K}\sum_\mu \xi_{i_1}^\mu\ldots \xi_{i_P}^\mu=\frac{1}{\sqrt{K}}\lambda_{i_1\ldots i_P}, \quad\quad\quad \lambda_{i_1\ldots i_P}\sim {\mathcal N}(0,1)
$$
and
$$
\frac{1}{K}\sum_\mu \lambda_{i_1\ldots i_P}^\mu\xi_{i_1}^\mu\ldots \xi_{i_P}^\mu=\frac{1}{\sqrt{K}}\lambda_{i_1\ldots i_P}, \quad\quad\quad \lambda_{i_1\ldots i_P}\sim {\mathcal N}(0,1).
$$
Thus the partition function of the unsupervised DAM model reads as 
\begin{align} \label{eq:vera_Z}
    \mathcal{Z}^{(P)}_{N} (\bm \eta^1,\bm\lambda; J) =&  \lim_{J \rightarrow 0} \sum_{\bm \sigma} \exp \left[ J \sum_{i=1}^N \xi_i^1 \si + \dfrac{\beta '}{2\R^{P/2}  M\,N^{P-1}}\SOMMA{a=1}{M}\left(\SOMMA{i}{N}\eta_{_{i}}^{1,a}\sigma_{_{i}}\right)^{^P} \right. \notag 
    \\
    &\left.+\frac{\alpha}{\sqrt{K}}\dfrac{\beta 'P!(1+\sqrt{\rho_P})}{2(1+\rho)^{P/2}}\left(\SOMMA{{i_1 <\cdots < i_P}}{N,\cdots,N}\lambda_{i_1,\hdots,i_P}\:\sigma_{_{i_1}}\cdots\sigma_{_{i_{_{P}}}}\right)\right].
\end{align}

\par\medskip
To make analytical progress, it is convenient to introduce the order parameters of the model.
\begin{definition}
The order parameters of the unsupervised DAM model are the Mattis magnetization $m_1$
of the archetype $\bm\xi^1$, the Mattis magnetizations $n_{a,1}$ of each example $a=1\ldots M$
of the archetype $\bm\xi^1$
and the two-replica overlaps $q_{lm}$, which quantifies the correlations between the variables 
${\bm \sigma}$ in two copies $l$ and $m$ of the system, sharing the same disorder (i.e. two {\em replicas}): 
  \begin{eqnarray}
\label{eq:def_m}
         m_{1}(\bm \sigma)&:=&\dfrac{1}{N}\SOMMA{i=1}{N}\xi_i^{1}\sigma_i \\
         \label{eq:def_n}
         n_{1,a} (\bm \sigma) &:=&\dfrac{r}{\R}\dfrac{1}{N}\SOMMA{i=1}{N}\eta_i^{1\,,a}\sigma_i,\\
         \label{eq:def_q}
         q_{lm}(\bm{\sigma}^{(l)},\bm{\sigma}^{(m)})&:=&\dfrac{1}{N}\SOMMA{i=1}{N}\sigma_i^{(l)}\sigma_i^{(m)}\,.
         \label{eq:def_p}
\end{eqnarray}
\end{definition}
In the next subsection we calculate the free energy of the system in the thermodynamic limit using Guerra's interpolation technique, assuming one step of replica symmetry breaking (1RSB). 


\subsection{1RSB analysis via Guerra's interpolation technique}\label{GuerraUnsup}
The key idea of Guerra's interpolation method is to define an auxiliary free energy, $\mathcal{F}^{(P)}(t)$, which is function of a parameter $t \in [0,1]$ that interpolates between the free energy of the original DAM model (obtained at $t=1$) and the free energy of an exactly solvable one-body model (obtained at $t=0$), whose effective fields mimic those 
of the original model. As the direct calculation of $\mathcal{F}^{(P)}(t=1)$ is cumbersome, this is obtained by using the fundamental theorem of calculus, namely we first evaluate $\mathcal{F}^{(P)}(t=0)$ and then we obtain $\mathcal{F}^{(P)}(t=1)$ as 
\begin{equation}
\mathcal{F}^{(P)}(t=1) = \mathcal{F}^{(P)}(t=0) + \int_0^1 ds\,\frac{d\mathcal{F}^{(P)}}{dt} \vert_{t=s}.
\label{eq:F_T_Calculus}
\end{equation}
To perform the analysis within the first step of replica symmetry breaking, we make 
the standard $1$RSB assumption \cite{MPV}, stated below:
\begin{assumption} \label{def:HM_RSBPspin}
In $1$RSB assumption,
the distribution of the two-replica overlap $q_{12}$, 
in the thermodynamic limit, displays two delta-peaks at the values $\bar{q}_1$ and $\bar{q}_2$, 
and the concentration on these two values is ruled by the parameter $\theta \in [0,1]$, namely
\begin{align}
\lim_{N \rightarrow + \infty} \mathbb{P}_N(q_{12})= \theta \delta (q_{12} - \bar{q}_1) + (1-\theta) \delta (q_{12} - \bar{q}_2), \label{limforq2Pspin}
\end{align}
where 
$$
\mathbb{P}_N(q_{12})=\sum_{\bm{\sigma}^{(1)},\bm{\sigma}^{(2)}}\sum_{\bm{\eta}}\mathbb{P}(\bm{\eta})\mathbb{P}_N(\bm{\sigma}^{(1)},\bm{\sigma}^{(2)}|\bm{\eta})
\delta (q_{12}-q_{12}(\bm{\sigma}^{(1)},\bm{\sigma}^{(2)}))
$$
and  $\mathbb{P}_N(\bm{\sigma}^{(1)},\bm{\sigma}^{(2)}|\bm{\eta})=
\mathcal{B}_N(\bm{\sigma}_1|\bm{\eta})\mathcal{B}_N(\bm{\sigma}_2|\bm{\eta})$ as replicas are conditionally independent, given the disorder 
$\bm{\eta}$.
\\[5mm]
The Mattis magnetizations $m_1$ and $n_{1,a}$ are assumed to be self-averaging at their equilibrium values $\bar{m}$ and $\bar{n}$, respectively, namely
\begin{align}
\lim_{N \rightarrow + \infty} \mathbb{P}_N(m_1) &= \delta (m_1 - \bar{m}),
\label{limformPspin} \\
\lim_{N \rightarrow + \infty} \mathbb{P}_N(n_{1,a}) &= \delta (n_{1,a} - \bar{n}).
\label{limfornPspin}
\end{align}
where $\mathbb{P}_N(m_1)=\mathbb{E}_{\bm{\eta}}\sum_{\bm{\sigma}}\mathcal{B}_N(\bm{\sigma}|\bm{\eta})
\delta (m_1-m_1(\bm{\sigma}))
$ and $\mathbb{P}_N(n_{1,a})=\mathbb{E}_{\bm{\eta}}\sum_{\bm{\sigma}}\mathcal{B}_N(\bm{\sigma}|\bm{\eta})
\delta (n_{1,a}-n_{1,a}(\bm{\sigma}))$, with $\mathbb{E}_{\bm{\eta}}f(\bm{\eta})=\sum_{\bm{\eta}} f(\bm{\eta}) \mathbb P({\bm{\eta}})$.
\end{assumption}
\begin{definition}
\label{def:part_Interpolante_RS}
Given the interpolating parameter $t \in [0,1]$, the constant $A_1,\ A_2, \ \psi \in \mathbb{R}$ to be fixed later on, and the i.i.d. standard Gaussian variables $Y_i^{(b)} \sim \mathcal{N}(0,1)$ for $i=1, \hdots , N$ and $b=1, 2$ (that must be averaged over as explained below), the Guerra's $1$RSB  interpolating partition function for the unsupervised DAM model is given by 
\begin{eqnarray}
     \mathcal{Z}^{(P)}_2(\bm\eta^1,\bm\lambda, \bm{Y}; J, t)  &\coloneqq& \sums \mathcal B^{(P)}_2 (\boldsymbol \sigma|\bm\eta^1,\bm \lambda, \bm{Y}; J, t)
     \label{eq:B2}
     \\     
     &=& \sums \exp{\Bigg[}J \sum_{i=1}^N \xi_i^1 \si+\dfrac{t \beta 'N}{2M}\left(1+\rho \right)^{P/2}\SOMMA{a=1}{M}n_{1,a}^{^P}(\bm\sigma) +\psi(1-t)N\SOMMA{a=1}{M}n_{1,a}(\bm\sigma)
     \nonumber
             \\
        &+&\sqrt{1-t}\SOMMA{b=1}{2}\left(A_b\SOMMA{i=1}{N}Y_i^{(b)}\sigma_i\right)+\sqrt{t}\dfrac{\beta 'P!\alpha(1+\sqrt{\rho_P})}{2(1+\rho)^{P/2} \sqrt{K}}\SOMMA{{i_1 <\cdots < i_P}}{N,\cdots,N}\lambda_{i_1,\hdots,i_P}\sigma_{_{i_1}}\cdots\sigma_{_{i_{_{P}}}}\Bigg].
        \nonumber\\
     \label{def:partfunct_GuerraRS}
\end{eqnarray}
The index $2$ on the LHS stands for the number of vectors ${\bm{Y}}^{(b)}$ that must be averaged over. Their number is equal to $k\!+\!1$ where $k$ is the number of steps of replica-symmetry breaking (here $k=1$). 

The average over the generalised Boltzmann factor associated to the interpolating partition function $\mathcal{Z}^{(P)}_2(\bm{\eta},\boldsymbol{\eta}, \bm{Y}; J, t)$, can be defined as 
\beq
\label{omegaNKM}
	\omega_{t} (\cdot) \coloneqq  \frac{1}{\mathcal Z^{(P)}_2( \bm{\eta}^1, \bm{\lambda}, \bm{Y};J, t )} \, \sum_{\boldsymbol \sigma}~ \cdot ~   \mathcal B_2^{(P)} 
 (\boldsymbol{\sigma} |\bm\eta^1, \bm{\lambda}, \bm{Y};J,  t ).
	\eeq

\begin{remark}
\label{r:2}
We note that for $t=1$, \eqref{def:partfunct_GuerraRS} recovers the original partition function 
\eqref{eq:vera_Z}, whereas for $t=0$ it reduces to the partition function of a system of $N$ non-interacting neurons, 
described by the Hamiltonian 
$-\b \mathcal{H}_{N}^{(1)}(\bm \sigma \vert \bm{\eta}, {\bf Y} ; J)=\sum_i h_i \sigma_i$, with local fields
$h_i=J\xi_i^1+\psi r\R^{-1}\sum_{a=1}^M \eta_i^{1,a}+\sum_{b=1}^2 A_bJ_i^{(b)}$, that is readily evaluated. The parameters of the one-body model $\psi$, $A_1$ and $A_2$ must be chosen 
in such a way 
that the $t$-dependent terms in  $d\mathcal{F}^{(P)}/dt$ cancel out \new{in the thermodynamic limit, under the 1RSB assumption.}
\end{remark}

In what follows, we will average over the 
fields ${\bm{Y}}^{(b)}$ for $b=1,2$ recursively, as explained by Guerra in \cite{guerra_broken} and in the statements below. To this purpose, we define
\begin{align}
\label{eqn:Z1Pspin}
\mathcal Z_1^{(P)}(\bm\eta^1,\bm \lambda,\bm{Y}^{(1)};  J, t)  \coloneqq& \left [\mathbb{E}_{\bm{Y}^{(2)}}  \mathcal Z_2^{(P)}(\bm\eta^1,\bm\lambda,\bm Y;  J, t) ^\theta \right ]^{1/\theta} \\
\label{eqn:Z0_1RSBPspin}
\mathcal Z_0^{(P)}(\bm\eta^1,\bm \lambda; J, t)  \coloneqq&  \exp \left\{\mathbb{E}_{\bm{Y}^{(1)}} \left[ \ln \mathcal Z_1^{(P)}(\bm\eta^1,\bm \lambda, \bm{Y}^{(1)}; J, t)  \right ]\right\} \\ \label{dimenticanza}
\mathcal{Z}^{(P)}_{N} (\bm\eta^1,\bm \lambda; J, t) \coloneqq & \mathcal Z_0^{(P)}(\bm\eta^1,\bm \lambda; J, t) ,
\end{align}
where with $\mathbb E_{\bm{Y}^{(b)}}[.]$ we mean the average over the vectors ${\bm{Y}}^{(b)}$ for $b=1, 2$.
%
%
The interpolating quenched free energy related to the partition function (\ref{def:partfunct_GuerraRS}) is introduced as
\begin{equation}
-\b\mathcal{F}^{(P)}_{N}(J, t) \coloneqq \frac{1}{N} \mathbb{E}_0 \left[  \ln \mathcal{Z}^{(P)}_{N}(\bm\eta^1,\bm \lambda; J, t)  \right],
\label{hop_GuerraAction}
\end{equation}
where $\mathbb{E}_0[.]$ denotes the average over the variables $\lambda^{\mu}_{i_1,\hdots,i_P}$'s and $\eta_{i}^{1,a}$'s. 
%
In the thermodynamic limit, assuming the limit exists, we write
\begin{equation}
\mathcal{F}^{(P)}(J, t) \coloneqq \lim_{N \to \infty} \mathcal{F}^{(P)}_{N}(J,t).
\label{hop_GuerraAction_TDL}
\end{equation}
\end{definition}
\begin{remark}
We note that in the RS case, where $\mathcal Z_2^{(P)}(\bm\eta^1,\bm \lambda, \bm Y; J, t) \equiv \mathcal Z_1^{(P)}(\bm\eta^1,\bm \lambda, \bm{Y}^{(1)}; J, t)$ and $\bm{Y}^{(2)}={0}$,
we have
\begin{equation}
-\b \mathcal{F}^{(P)}(J, t) \coloneqq \lim_{N \to \infty} \frac{1}{N} \mathbb{E}_0 \mathbb{E}_{\bm{Y}^{(1)}}\left[  \ln \mathcal{Z}^{(P)}_{1}(\bm\eta^1,\bm\lambda, \bm{Y}^{(1)}; J, t)  \right],
\end{equation}
%
hence the average $\mathbb{E}_{\bm Y^{(1)}}$ acts on the same level as $\mathbb{E}_0$ i.e. outside the logarithm. When moving from the RS to the $1$RSB scenario, an extra average is introduced, $\mathbb{E}_{\bm Y^{(2)}}$, which acts on a different level 
than $\mathbb{E}_0$ and $\mathbb{E}_{\bm Y^{(1)}}$ i.e. {\it inside} the logarithm. 
This reflects the presence of hierarchical valleys in the free energy landscape \cite{Rammal} which leads to the well-known multi-scale thermalization in spin glasses. The internal average $\mathbb{E}_{Y^{(2)}}$ is related to thermalization within a valley at the lower level of the (two-level) hierarchy, whereas the external averages account for thermalization 
across valleys (i.e. at the higher level of the hierarchy). The structure of the hierarchy is captured by the parameter $\theta$, which controls the amplitudes of the valleys at the lower level of the hierarchy and can be related to {\em effective temperatures} \cite{Leticia}, as discussed in \cite{VanMourik} for Parisi's replica approach and in \cite{Mingione} for Guerra's interpolation techniques. 
Note that the distinction between the two levels of the hierarchy is not made for $\psi$ which accounts for the signal contribution and is kept RS (as usually done for the magnetization in spin-glass models).

\end{remark}

Now, following Guerra's prescription \cite{guerra_broken}, 
given two copies (or replicas) of the system, we define the following averages, corresponding to thermalization within the two different 
levels of the hierarchy
\begin{align}
    &\langle \cdot   \rangle_{t,1}  \coloneqq \mathbb{E}_0\mathbb{E}_{\bm Y^{(1)}}[\mathbb{E}_{\bm Y^{(2)}} \mathcal{W}_{2,t} \omega_{t}( \cdot)]^2, 
    \label{eq:media1} \\
    &\langle \cdot   \rangle_{t,2}  \coloneqq \mathbb{E}_0\mathbb{E}_{\bm Y^{(1)}}\mathbb{E}_{\bm Y^{(2)}} \mathcal{W}_{2,t} [\omega_{t}(\cdot) ]^2,\label{eq:media2}
\end{align}    
where 
\begin{equation}
\label{eq:W2}
    \mathcal{W}_{2,t} \coloneqq \dfrac{\mathcal{Z}_{2}^\theta(\bm{\eta}^1,\bm\lambda, \bm Y;J,t)}{\mathbb{E}_{Y^{(2)}} \mathcal{Z}^\theta_{2}(\bm{\eta}^1,\bm\lambda, \bm Y;J,t)}.
\end{equation}

At this point, we are able to state our second assumption, that is at each level $a=1,2$ of the hierarchy, the two-replica overlap 
$q_{12}(\bm{\sigma}^{(1)},\bm{\sigma}^{(2)})$ self-averages around the value $\bar{q}_a$ of the corresponding peak in the overlap distribution $\mathbb{P}_N(q)$, as given in Assumption \ref{def:HM_RSBPspin}:
\begin{assumption}
\label{eq:seconda}
For any $t\in[0;1]$
\begin{eqnarray}
\langle [q_{12}(\bm\sigma^{(1)},\bm\sigma^{(2)}) -\bar{q}_a]^k\rangle_{t,a} &\xrightarrow[]{N\to\infty}&0\;\;\; \mathrm{for}\; a=1,2\;\;\mathrm{and}\;k\geq 2
\notag
\end{eqnarray}
\end{assumption}

Finally, we provide the explicit expression of the quenched free energy in terms of the control
parameters in the next

\begin{proposition}
\label{P_quenched}
In the thermodynamic limit $N\to\infty$, within the $1$RSB Assumption and 
under the Assumption \ref{eq:seconda}, the quenched free energy for the unsupervised DAM model, reads as
\begin{equation}
\label{eq:pressure_GuerraRS_finale}
\begin{array}{lll}
     -\b\mathcal{F}^{(P)}
    (J=0) =&\ \dfrac{1}{\theta}\mathbb{E}_{\bm\xi^1}\mathbb{E}_{(\bm\eta^1|\bm\xi^1)}\mathbb{E}_{Y^{(1)}}\ln\mathbb{E}_{Y^{(2)}} \left\{ \cosh^\theta\left[
    \beta '\dfrac{P}{2}\n^{P-1}(1+\rho)^{P/2-1} \etaM \right. \right.
     \\\\
     &\left. \left. +Y^{(1)} \beta '\sqrt{\gamma
     \dfrac{P}{2}\dfrac{(1+\sqrt{\rho_P})^2}{(1+\rho)^{P} } \q_1^{^{P-1}}} +Y^{(2)} \beta '\sqrt{\gamma
     \dfrac{P}{2}\dfrac{(1+\sqrt{\rho_P})^2}{(1+\rho)^{P} } \left(\q_2^{P-1}  - \q_1 ^{^{P-1}}\right)}\right]\right\} 
     \\\\
         &-\b \dfrac{1}{2}(P-1)(1+\rho)^{P/2}\n^P + \dfrac{{\b}^2}{4}\gamma\dfrac{(1+\sqrt{\rho_P})^2  }{ (1+\rho)^{P}}\left[1-P\q_2^{P-1}+(P-1)\q_2^P\right]
         \\\\
         &-\dfrac{{\b}^2}{4}\gamma\dfrac{(1+\sqrt{\rho_P})^2}{ (1+\rho)^{P}}(P-1)\theta(\q_2^P-\q_1^P)+\ln 2.
\end{array}
\end{equation}
where 
\begin{equation}
\etaM:=\dfrac{1}{rM}\sum_{a=1}^M\eta^{1,a},
\label{eq:etaM}
\end{equation}
\new{$\mathbb{E}_{(\bm{\eta}^1|\bm \xi^1)}=\prod\limits_{a=1}^M \mathbb{E}_{(\eta^{1,a}|\xi^1)}$ and 
$\mathbb{E}_{\bm \xi^1}$ and $\mathbb{E}_{(\eta^{1,a}|\xi^1)}$
denote the expectations over the Bernoulli distributions \eqref{eq:xi} and
\eqref{eq:Bernoulli} respectively}. In the above, $\bar n$, $\q_1, \ \q_2$ fulfill the following self-consistency equations
\begin{equation}
    \begin{array}{lll}
         \n=\dfrac{1}{1+\rho}\mathbb{E}_{\bm\xi^1}\mathbb{E}_{(\bm\eta^1|\bm\xi^1)} \mathbb{E}_{Y^{(1)}} \left[\dfrac{\mathbb{E}_{Y^{(2)}} \cosh^\theta g(\beta, \gamma, \bm Y, \bm\eta^1) \tanh g(\beta, \gamma, \bm Y, \bm\eta^1)}{\mathbb{E}_{Y^{(2)}} \cosh^\theta g(\beta, \gamma, \bm Y, \bm\eta^1)} \etaM \right],
         \\\\
         \q_1= \mathbb{E}_{\bm\xi^1}\mathbb{E}_{(\bm\eta^1|\bm\xi^1)} \mathbb{E}_{Y^{(1)}} \left[\dfrac{\mathbb{E}_{Y^{(2)}} \cosh^\theta g(\beta, \gamma, \bm Y, \bm\eta^1) \tanh g(\beta, \gamma, \bm Y, \bm\eta^1)}{\mathbb{E}_{Y^{(2)}} \cosh^\theta g(\beta, \gamma, \bm Y, \bm\eta^1)} \right]^2,
         \\\\
         \q_2 = \mathbb{E}_{\bm\xi^1}\mathbb{E}_{(\bm\eta^1|\bm\xi^1)} \mathbb{E}_{Y^{(1)}} \left[\dfrac{\mathbb{E}_{Y^{(2)}} \cosh^\theta g(\beta, \gamma, \bm Y, \bm\eta^1) \tanh^2 g(\beta, \gamma, \bm Y, \bm\eta^1)}{\mathbb{E}_{Y^{(2)}} \cosh^\theta g(\beta, \gamma, \bm Y, \bm\eta^1)} \right],
    \end{array}
    \label{eq:High_store_self_n_q}
\end{equation}
with 
\begin{align}
    g(\beta, \gamma, \bm Y, \bm\eta^1) =& \beta '\dfrac{P}{2}\n^{P-1}(1+\rho)^{P/2-1} \etaM+Y^{(1)} \beta '\sqrt{\gamma\dfrac{(1+\sqrt{\rho_P})^2}{(1+\rho)^{P}}  \dfrac{P}{2}\q_1^{^{P-1}}} \notag \\
    &+Y^{(2)} \beta '\sqrt{
    \gamma\dfrac{(1+\sqrt{\rho_P})^2}{(1+\rho)^{P}}\dfrac{P}{2} \left(\q_2^{^{P-1}} -  \q_1 ^{^{P-1}}\right)}
    \label{eq:g}
\end{align}
and $\b= 2\beta/P!$.
Furthermore, as $\bar{m}= -\b\nabla_J \mathcal{F}^{(P)}(J)|_{J=0}$, we have
\begin{align}
\label{eq:High_store_self_m}
    \m= \mathbb{E}_{\bm\xi^1}\mathbb{E}_{(\bm\eta^1|\bm\xi^1)}\mathbb{E}_{Y^{(1)}} \left[\dfrac{\mathbb{E}_{Y^{(2)}} \cosh^\theta g(\beta, \gamma, \bm Y, \bm\eta^1) \tanh g(\beta, \gamma, \bm Y, \bm\eta^1)}{\mathbb{E}_{Y^{(2)}} \cosh^\theta g(\beta, \gamma, \bm Y, \bm\eta^1)}\xi^1\right] .
    \end{align}
\end{proposition}

In order to prove the aforementioned proposition, we need to premise the following 
\begin{lemma} 
\label{lemma:der}
In the thermodynamic limit, the $t$ derivative of the interpolating quenched free energy \eqref{hop_GuerraAction} under the 1RSB assumption and under Assumption \ref{eq:seconda} is given by 

\begin{equation}
    \begin{array}{lll}
         \dfrac{d \mathcal{F}^{(P)}(J,t)}{d t} = &  \dfrac{1}{2}(P-1)(1+\rho)^{P/2}\n^P - \dfrac{\b}{4}\gamma\dfrac{(1+\sqrt{\rho_P})^2}{ (1+\rho)^{P}}\left[1-P\q_2^{P-1}+(P-1)\q_2^P\right]
         \\\\
         &+\dfrac{\b}{4}\gamma\dfrac{(1+\sqrt{\rho_P})^2}{ (1+\rho)^{P}}(P-1)\theta(\q_2^P-\q_1^P).
    \end{array}
    \label{eq:streaming_RS_Guerra}
\end{equation}

\end{lemma}

The proof of this Lemma is shown in Appendix \ref{app:lemma}. Now let us prove Proposition \ref{P_quenched}:
\begin{proof}
Exploiting the fundamental theorem of calculus, we can relate the free energy of the original model $\mathcal{F}^{(P)}(J,t=1)$ and the one stemming from the one body terms $\mathcal{F}^{(P)}(J,t=0)$ via \eqref{eq:F_T_Calculus}.
\new{Since, 
the $t$-derivative \eqref{eq:streaming_RS_Guerra} does not depend on $t$, 
all we need is to add the expression above to the free energy $\mathcal{F}^{(P)}(J,t=0)$, which only contains one body terms}. Let us start with the computation of the latter at finite size $N$:
\begin{align}
     -\b\mathcal{F}_{N}^{(P)}(J,t=0)=& \dfrac{1}{\theta N} \mathbb{E}_{\bm\xi^1}\mathbb{E}_{(\bm\eta^1|\bm\xi^1)} \mathbb{E}_{Y^{(1)}} \ln \mathbb{E}_{Y^{(2)}}  \sum_{\bm \sigma}  \exp \left[-\beta \mathcal{H}_{N}^{(1)}(\bm \sigma \vert \bm\eta^1, {\bf Y}; J)\right]
     \notag \\
     =& \ln 2 + \dfrac{1}{\theta} \mathbb{E}_{\bm\xi^1}\mathbb{E}_{(\bm\eta^1|\bm\xi^1)} \mathbb{E}_{Y^{(1)}} \ln \mathbb{E}_{Y^{(2)}} \cosh^{\theta} \left(J\xi^1 +  \dfrac{\psi}{(1+\rho)} \hat\eta_{M} + Y^{(1)} A_1 + Y^{(2)}A_2\right),
\end{align}
with the definition of $-\beta \mathcal{H}_{N}^{(1)}(\bm \sigma \vert \bm \eta; J)$ given in Remark \ref{r:2}.

\new{Upon setting the constants to the values \eqref{eq:values} stated in the proof of Lemma \ref{lemma:der}, we have}
\begin{align}
     -\b\mathcal{F}_{N}^{(P)}(J,t=0)=&\ln 2 + \dfrac{1}{\theta} \mathbb{E}_{\bm\xi^1}\mathbb{E}_{(\bm\eta^1|\bm\xi^1)} \mathbb{E}_{Y^{(1)}} \ln \mathbb{E}_{Y^{(2)}} \cosh^\theta g(\b, K, N, \bm Y),
     \label{eq:Aproof}
\end{align}

where $g(\beta, K,N, \bm Y)$ reads as 
\begin{align}
        g(\beta, K, N , \bm Y) =& J \xi^1+\beta '\dfrac{P}{2}\n^{P-1}(1+\rho)^{P/2-1} \etaM\notag \\
        &+Y^{(1)} \beta '\sqrt{\dfrac{P(1+\sqrt{\rho_P})^2 }{2(1+\rho)^{P}} \dfrac{P!K}{2N^{P-1}} \q_1^{^{P-1}}} +Y^{(2)} \beta '\sqrt{
    \dfrac{P(1+\sqrt{\rho_P})^2 }{2(1+\rho)^{P}} \dfrac{P!K}{2N^{P-1}}\left(\q_2^{^{P-1}} -  \q_1 ^{^{P-1}}\right)}.
    \label{eq:gproof}
\end{align}

\new{Recalling that $\lim\limits_{N\to \infty}K/N^{P-1} =2\gamma/P!$, 
taking the thermodynamic limit of the above equations and inserting them in the fundamental theorem of calculus \eqref{eq:F_T_Calculus}, we reach \eqref{eq:pressure_GuerraRS_finale}}.
%
\new{Finally, by maximizing \eqref{eq:pressure_GuerraRS_finale}
w.r.t. to the order parameters $\bar{n}, \bar{q}_1, \bar{q}_2$, we obtain the self-consistency equations \eqref{eq:High_store_self_n_q}, hence we reach the thesis.} 
\end{proof}


%

\subsection{1RSB analysis via Parisi's replica trick}\label{ParisiUnsup}

In this Section, we derive the expression of the quenched free energy of the unsupervised DAM model provided in Proposition \ref{P_quenched} by using the replica method \cite{SK1975,MPV}, at the first step of replica-symmetry breaking \cite{GiorgioRSB, MPV}.
The core of this approach consists in writing the quenched free energy, as defined in \eqref{PressureDef_unsup}, as
\begin{align}
    -\b \mathcal{F}_N^{(P)}(J)=  \dfrac{1}{N} \lim_{n\to 0} \dfrac{\mathbb{E}\mathcal{Z}_N^n(J)-1}{n},
\end{align}
\new{with $\b = 2 \beta/P!$ and where we have used the shorthand  $\mathcal{Z}_N(J)=\mathcal{Z}_N^{(P)}(\bm{\eta};J)$ for the partition function, defined as
\begin{align}
   \mathcal{Z}_N(J) &= \sum_{\bm{\sigma}} \exp\left( -\b  \mathcal{H}^{(P)}_N(\bm{\sigma}| \bm{\eta}) + J \SOMMA{i=1}{N} \xi_i^1\sigma_i\right)\,,
   \label{eq:new_part_rt_unsup}
\end{align}
$\mathbb{E}[.]$ denotes, as in \eqref{PressureDef_unsup}, the average over the quenched disorder $\bm{\eta}$ and in accordance with the explanation provided in Sec. \ref{sec:unsup_gen}, the inclusion of the last term in the round brackets of \eqref{eq:new_part_rt_unsup} is intended to make the free energy a moment-generating function for the Mattis magnetization $m_1(\bm\sigma)$.}

For integer $n$ the function $\mathcal{Z}_N^n$ is the product of a system of $n$ identical replicas of the original system
\begin{align}
    \mathbb{E}\mathcal{Z}_N^n(J) &= \mathbb{E} \sum_{\bm{\sigma}^1 \ldots \bm{\sigma}^n} \exp\left( -\b \sum_{a=1}^N \mathcal{H}^{(P)}_N(\bm{\sigma}^{(a)}| \bm{\eta}) + J \SOMMA{a=1}{n}\SOMMA{i=1}{N} \xi_i^1\sigma_i^{(a)}\right) 
\end{align}
namely 
\begin{align}
     \mathbb{E}\mathcal{Z}_N^n(J) &= \mathbb{E} \sum_{\bm{\sigma}^1 \ldots \bm{\sigma}^n} \exp\left[ \dfrac{\b P!}{2\R^{P/2}\,M\,N^{P-1}}\SOMMA{a=1}{n}\SOMMA{\mu=1}{K}\SOMMA{A=1}{M}\left(\SOMMA{i_1 <\cdots < i_P}{N,\hdots,N}\eta^{\mu\,,A}_i\cdots\eta^{\mu\,,A}_{i_P}\sigma^{(a)}_{_{i_1}}\cdots\sigma^{(a)}_{_{i_P}}\right)+ J \SOMMA{a, i}{n,N}\xi_i^1\sigma_i^{(a)}\right]\,.\label{eq:Z_rt_unsup}
\end{align}
Splitting, as before, the signal term ($\mu=1$) from the remaining terms $\mu>1$, which act as a quenched noise on the learning and retrieval of pattern $\bm{\xi}^1$, we can write
\begin{align}
    \mathbb{E}\mathcal{Z}_N^n(J) =& \sum_{\bm{\sigma}^1\ldots \bm{\sigma}^n} \mathbb{E}_{\bm \xi^1}\mathbb{E}_{(\bm \eta^{1}|\bm\xi^1)}  \exp\left[ \dfrac{\b (1+\rho)^{P/2} }{2  MN} \sum_{a=1}^n\sum_{A=1}^M\left( \dfrac{r}{N \R}\sum_i \eta^{1\,,A}_i \sigma_i^{(a)} \right)^P + J \SOMMA{a, i}{n,N}\xi_i^1\sigma_i^{(a)}\right] \cdot \notag 
    \\
    &\mathbb{E}_{\bm \xi^{\mu>1}}\mathbb{E}_{(\bm \eta^{\mu>1}|\bm\xi^{\mu>1})}  \exp \left[\dfrac{\b P!}{2 M\R^{P/2}N^{P-1}} \sum_{\mu>1, a,A}\sum_{i_1 <\cdots < i_P} \eta^{\mu\,,A}_{i_1}\hdots \eta^{\mu\,,A}_{i_{P}} \sigma_{i_1}^{(a)}\hdots \sigma_{i_{P}}^{(a)}\right].
    \label{eq:first}
\end{align}
\new{where $\mathbb{E}_{\bm{\xi}^\mu}[.]=\prod_i \mathbb{E}_{\xi_i^\mu}[.]$ denotes the average over the pattern $\bm{\xi}^\mu$, whose entries $\xi_i^\mu$ are drawn from the distribution \eqref{eq:xi} and $\mathbb{E}_{(\bm{\eta}^\mu|\bm{\xi}^\mu)}=\prod_{i,a}\mathbb{E}_{(\eta_i^{\mu,a}|\xi_i^\mu)}$ with 
$\mathbb{E}_{\eta_i^{\mu,A}}$ denoting the 
expectation over the conditional distribution 
\eqref{eq:Bernoulli}.}
As before, exploiting the CLT to rewrite the last term in \eqref{eq:first} 
via \eqref{eq:con_V}, we obtain
\begin{align}
    \mathbb{E}\mathcal{Z}_N^n(J) =& \sum_{\bm{\sigma}^1\ldots \bm{\sigma}^n} \mathbb{E}_{\bm \xi^1}\mathbb{E}_{(\bm \eta^{1}|\bm\xi^1)}  \exp\left[ \dfrac{\b (1+\rho)^{P/2}N }{2  M} \sum_{a=1}^n\sum_{A=1}^M\left( \dfrac{r}{N \R}\sum_i \eta^{1\,,A}_i \sigma_i^{(a)} \right)^P + J \SOMMA{a, i}{n,N}\xi_i^1\sigma_i^{(a)}\right] \cdot \notag 
    \\
    &\mathbb{E}_{\bm \lambda}  \exp \left[\dfrac{\b P! \alpha(1+\sqrt{\rho_P})}{2 (1+\rho)^{P/2}\sqrt{K}} \sum_{a}\sum_{i_1 <\cdots < i_P}  \lambda_{i_1,\hdots,i_{P}} \sigma_{i_1}^{(a)}\hdots \sigma_{i_{P}}^{(a)}\right]\notag 
    \\
    =& \sum_{\bm{\sigma}^1\ldots \bm{\sigma}^n} \mathbb{E}_{\bm \xi^1}\mathbb{E}_{(\bm \eta^{1}|\bm\xi^1)} \mathcal{Z}_{signal}^n(J) \mathbb{E}_{\bm \lambda} \mathcal{Z}_{noise}^n,
    \label{eq:replicated-Z}
\end{align}
\new{
Treating the two terms $\mathbb{E}_{\bm \xi^{1}}\mathbb{E}_{(\bm \eta^{1}|\bm\xi^{\1})} \mathcal{Z}_{signal}^n(J)$ and $\mathbb{E}_{\bm \lambda} \mathcal{Z}_{noise}^n$ separately, 
we insert in the former the identity
$$
1=\prod\limits_{a,A=1,1}^{n,M}\int d n_{1,A}^{(a)}\, \delta\left(n_{1,A}^{(a)}  -  n_{1,A}^{(a)}(\bm{\sigma})\right)
$$
and use the Fourier representation of the Dirac delta, obtaining
\begin{align}
 \mathbb{E}_{\bm \xi^{1}}\mathbb{E}_{(\bm \eta^{1}|\bm\xi^{1})} \mathcal{Z}_{signal}^n(J) =& \mathbb{E}_{\bm \xi^{1}}\mathbb{E}_{(\bm \eta^{1}|\bm\xi^{1})} \int \prod_{a,A} d n_{1,A}^{(a)} \prod_{a,A} \frac{dz_{1,A}^{(a)}}{2\pi} \exp \left[ \dfrac{\b N (1+\rho)^{P/2}}{2M} \sum_{A,a} (n_{1,A}^{(a)})^P \right. \notag 
    \\
    &\left.+i \sum_{a,A} \left(n_{1,A}^{(a)}  - \dfrac{r}{\R N} \sum_{i} \eta_i^{1,A}  \sigma_i^{(a)} \right)z_{1,A}^{(a)} + J \SOMMA{a, i}{n,N}\xi_i^1\sigma_i^{(a)}\right].
    \label{eq:rtsig}
    \end{align}
For the noise term, performing first the expectation over $\bm\lambda$ 
\begin{align}
    \mathbb{E}_{\bm \lambda} \mathcal{Z}_{noise}^n =& 
    \int \prod_{\bm i} \dfrac{d\lambda_{i_1,\hdots,i_P}}{\sqrt{2\pi}}  \exp \left[-\dfrac{(\lambda_{i_1,\hdots,i_P})^2}{2}+\dfrac{\b  \alpha(1+\sqrt{\rho_P})}{2 (1+\rho)^{P/2}\sqrt{K}}  P!\SOMMA{a}{n}\SOMMA{i_1<\hdots<i_P}{N,\hdots,N}\lambda_{i_1,\hdots,i_P}\sigma_{i_1}^{(a)}\cdots\sigma_{i_P}^{(a)}   \right] \notag 
    \\
    =& \exp \left[\dfrac{1}{2}\left(\dfrac{\b (1+\sqrt{\rho_P})}{2 (1+\rho)^{P/2}N^{P-1}}\right)^2  K(P!)^2 \SOMMA{i_1<\hdots<i_P}{N,\hdots,N}\SOMMA{a,b}{n,n}\sigma_{i_1}^{(a)}\sigma_{i_1}^{(b)}\cdots\sigma_{i_P}^{(a)}\sigma_{i_P}^{(b)}   \right] \notag
    \\
    =& \exp \left[\dfrac{1}{2}\left(\dfrac{\b (1+\sqrt{\rho_P})}{2 (1+\rho)^{P/2}N^{P/2-1}}\right)^2 K  P! \SOMMA{a,b}{n,n}\left(\dfrac{1}{N}\SOMMA{i}{N}\sigma_{i_1}^{(a)}\sigma_{i_1}^{(b)}\right)^P   \:\right] \notag
    \\
    \end{align}
inserting 
$$
1=\prod_{a,b}\int dq_{ab}\, \delta(q_{ab}-q_{ab}(\bm{\sigma}))
$$
and using the Fourier representation of the Dirac delta, we have 
\begin{align}
    \mathbb{E}_{\bm \lambda} \mathcal{Z}_{noise}^n =& 
    \int \prod_{a,b} dq_{ab} \int \prod_{a,b} \dfrac{dp_{ab}}{2\pi} \exp \left[\dfrac{1}{2}\left(\dfrac{\b  (1+\sqrt{\rho_P})^2}{2 (1+\rho)^{P/2}N^{P/2-1}} \right)^2 K P!\SOMMA{a,b}{n,n}q_{ab}^P   \right. \notag \\
    &\left. +i \sum_{a,b} \left( q_{ab} - \dfrac{1}{N}\sum_{i}\sigma_i^{(a)}\si^{(b)} \right)p_{ab} \right]
    \label{eq:rtnoise}
\end{align}
}
Inserting Eqs. \eqref{eq:rtsig} and \eqref{eq:rtnoise} in \eqref{eq:replicated-Z}, we obtain
\begin{align}\label{Laplace}
    &\mathbb{E}\mathcal{Z}_N^n(J) =  \int \prod_{a,A} d n_{1,A}^{(a)} \prod_{a,A} \dfrac{dz_{1,A}^{(a)}}{2\pi} \int \prod_{a,b} dq_{ab} \int \prod_{a,b} \dfrac{dp_{a,b}}{2\pi} \exp\left( -N A[\bm Q, \bm P, \bm Z, \bm N] \right) 
\end{align}
where
\begin{align}
    &A[\bm Q, \bm P, \bm Z, \bm N]= - \dfrac{i}{N} \sum_{a,b} p_{ab}q_{ab} -\dfrac{1}{4}\dfrac{\b\,^2  (1+\sqrt{\rho_P})^2}{ (1+\rho)^{P}} \gamma \SOMMA{a,b}{n,n}q_{ab}^P - \dfrac{\b (1+\rho)^{P/2}}{2M} \sum_{A,a} (n_{1,A}^{(a)})^P \notag \\
    & -\dfrac{i}{N}\sum_{a,A} n_{1,A}^{(a)} z_{1,A}^{(a)}
    -\mathbb{E}_{\bm \xi^{1}}\mathbb{E}_{(\bm \eta^{1}|\bm\xi^{1})} \log \left[\sum_{\bm \sigma} \exp\left( -\dfrac{i}{N} \sum_{a,b,i} p_{ab} \sigma^{(a)}\sigma^{(b)} - \dfrac{i r}{\R N} \sum_{a,A} \eta^{1,A} z_{1,A}^{(a)} \sigma^{(a)}+ J \SOMMA{a}{}\xi^1\sigma^{(a)}\right)\right].
    \label{eq:A_star}
\end{align}
and we have used the shorthand notation ${\bm Q}=\{q_{ab}\}_{a,b=1}^n$ and similarly for ${\bm P}$, 
${\bm N}$ and ${\bm Z}$.
\new{Next, we assume that the limit $n\to 0$  in 
\begin{align}
    -\b \mathcal{F}^{(P)}(\beta)= \lim_{N\to\infty} \dfrac{1}{N} \lim_{n\to 0} \dfrac{\mathbb{E}\mathcal{Z}_N^n-1}{n},
\end{align}
can be taken by analytic continuation and that the two limits $N\to\infty$ and $n\to 0$ can be interchanged (provided they exist), so that the integrals in \eqref{Laplace}
can be performed by steepest descent.} This leads to
\begin{align}
    -\b \mathcal{F}^{(P)}(\beta)=- \lim_{n\to 0} \dfrac{1}{n} A[\bm Q^\star, \bm P^\star, \bm Z^\star, \bm N ^\star] 
    \label{eq:trick_relation}
\end{align}
where $\bm Q^\star, \bm P^\star, \bm Z^\star, \bm N ^\star$, are solution of the saddle point equations 
\begin{eqnarray}
&\dfrac{\partial A}{\partial p_{ab}}=0&\quad\hence\quad 
q_{ab}^*=\langle \sigma^a \sigma^b\rangle_{\rm eff}
\label{eq:selfQ}
\\
&\dfrac{\partial A}{\partial z_{1,A}^{(a)}}=0&\quad\hence\quad 
(n_{1,A}^{(a)})^*=\dfrac{r}{\R}\SOMMA{A=1}{M}\mathbb{E}\langle \eta^{1,A}\sigma^a \rangle_{\rm eff}
\label{eq:selfm}
\\
&\dfrac{\partial A}{\partial q_{ab}}=0&\quad\hence\quad 
p_{ab}^*=\dfrac{i \b\,^2}{2} \dfrac{P}{2}\gamma \dfrac{(1+\sqrt{\rho_P})^2}{(1+\rho)}N (q_{ab}^*)^{P-1}
\label{eq:selfP}
\\
&\dfrac{\partial A}{\partial n_{1,A}^{(a)}}=0&\quad\hence\quad 
(z_{1,A}^{(a)})^*=i \b \dfrac{P}{2} \dfrac{(1+\rho)^{P/2}}{M} N [(n_{1,A}^{(a)})^*]^{P-1}
\label{eq:selfZ}
\end{eqnarray}
where the average $\langle \ldots \rangle_{\rm eff}$ is computed over the distribution $p_{\rm eff}(\bm{\sigma})=e^{-\b H_{\rm eff}(\bm{\sigma})}/Z_{\rm eff}$ where $H_{\rm eff}(\bm{\sigma})$ is equal to the terms in the round brackets in the second line of \eqref{eq:A_star}. 
\new{Inserting \eqref{eq:selfP} and \eqref{eq:selfZ} in $H_{\rm eff}(\bm{\sigma})$, we obtain 
\begin{equation}
\label{eq:Heff}
-\b H_{\rm eff}(\bm{\sigma})=\dfrac{{\b}^2\gamma}{2}\frac{P(1+\sqrt{\rho_P})^2}{2(1+\rho)^P} \sum_{a,b}  [q_{ab}^*]^{P-1} \sigma^{(a)}\sigma^{(b)}+ \b \dfrac{P}{2} (1+\rho)^{P/2-1}  \sum_{a}\dfrac{1}{rM}\sum_{A} \eta^{1,A}  [(n_{1,A}^{(a)})^*]^{P-1} \sigma^{(a)}
\end{equation}
and, substituting in  \eqref{eq:trick_relation}
\begin{align}
    -\b  \mathcal F^{(P)}(J)=&\lim_{n\to 0} \dfrac{1}{n} \left\{ \dfrac{{\b}^2(1+\sqrt{\rho_P})^2\gamma}{4(1+\rho)^P}  \sum_{a,b} (P-1) [q_{ab}^*]^{P} +\frac{\b (1+\rho)^{P/2} (P-1)}{2M}\sum_{a,A}   [(n_{1,A}^{(a)})^*]^{P}\right.\notag \\
    &    +\left.\mathbb{E}_{\bm \xi^{1}}\mathbb{E}_{(\bm \eta^{1}|\bm\xi^{1})} \log \left[ \sum_{\bm \sigma} \exp\left(-\b H_{\rm eff}+ J \SOMMA{a}{}\xi_i^1\sigma_i^{(a)}\right)\right]\right\}\,.
    \label{eq:rt_free}
\end{align}
where $q_{ab}^\star$ and $[n_{1,A}^{(a)}]^*$ must be determined from \eqref{eq:selfQ}, \eqref{eq:selfm}. 
Since $p_{\rm eff}(\bm{\sigma})$ depends on $\bm Q$,  $\bm N$ via \eqref{eq:Heff}, Eqs. \eqref{eq:selfQ} and \eqref{eq:selfm} denote a set of self-consistency equations.}
%
\new{Now, in order to proceed, we need to find the form of $\bQ$ and $\bm N$ in the limit $n\to 0$. We will make the $1$RSB ansatz}
:  
\begin{align}
    q_{ab}&=\delta_{ab} + (1-\delta_{ab})(\theta \q_1 +(1-\theta) \q_2), \label{eq:qabassump}\\
    n_{1,A}^{(a)}&= \n, 
\end{align}
\new{Using \eqref{eq:qabassump} to evaluate the first term in \eqref{eq:rt_free} and the first term in \eqref{eq:Heff}}, we get 
\begin{align}
    \sum_{ab} q_{ab}^P =&n \left(1+n \theta\q_1^P -\theta\q_1^P - \q_2^P + n \q_2^P + \theta\q_2^P  - n \theta\q_2^P \right)\label{eq:pabqab}\\
    \sum_{ab} q_{ab}^{P-1} \sigma^{(a)}\sigma^{(b)} =& \frac{1}{2} \left[ 2(1-\bar{q}_2^{P-1}) n
    + (\bar{q}_2^{P-1} - \bar{q}_1^{P-1}) \sum_{k=1}^{n/\theta}\left( \sum_{a=1}^\theta \sigma^{((k-1)\theta+a)}\right)^2+ \bar{q}_1^{P-1} \left( \sum_a \sigma^{(a)}\right)^2\right] \label{eq:pabsasb}.
\end{align}
Upon inserting in \eqref{eq:rt_free} we obtain 
\new{
\begin{align}
    &-\b  \mathcal F^{(P)}(J)= \dfrac{{\b}^2(1+\sqrt{\rho_P})^2\gamma}{4(1+\rho)^P}   [1 -  (P-1)\theta \left(\q_2^P-\q_1^P\right) +(P-1)\q_2^P] 
    -\frac{\b (1+\rho)^{P/2} (P-1)}{2}\bar{n}^{P}\notag \\
    &  -\frac{(\b)^2(1+\sqrt{\rho_P})^2\gamma P}{4(1+\rho)^P} \bar{q}_2^{P-1}  +\lim_{n\to 0}\frac{1}{n}\mathbb{E}_{\bm \xi^{1}}\mathbb{E}_{(\bm \eta^{1}|\bm\xi^{1})} \log \left[ \sum_{\bm \sigma} \exp\left(-\b H_{\rm eff}(\bm{\sigma})+ J \SOMMA{a}{}\xi^1\sigma^{(a)}\right)\right]\,
    \label{eq:rt_free-ans}
\end{align}
with 
\begin{eqnarray}
\label{eq:Heff-ans}
-\b H_{\rm eff}(\bm{\sigma})&=&{\b}^2\gamma\frac{P(1+\sqrt{\rho_P})^2}{4(1+\rho)^P}
\left[ 
    (\bar{q}_2^{P-1} - \bar{q}_1^{P-1}) \sum_{k=1}^{n/\theta}\left( \sum_{b=1}^\theta \sigma^{((k-1)\theta+b)}\right)^2+ \bar{q}_1^{P-1} \left( \sum_{a=1}^n \sigma^{(a)}\right)^2\right] 
\nonumber\\
&&+ \b \dfrac{P}{2}(1+\rho)^{P/2-1}  \hat{\eta}_M  \left(\sum_{a=1}^n\sigma^{(a)}\right) \bar{n}^{P-1}
\end{eqnarray}
}where we have used the definition \eqref{eq:etaM}.

Next, we apply a Gaussian transformation to the last term in \eqref{eq:rt_free-ans} to linearize the squared terms in the Hamiltonian \eqref{eq:Heff-ans}
\begin{align}
    &\lim_{n\to 0}\dfrac{1}{n}\ln \left[ \sum_{\bm \sigma} \exp\Big(-\b H_{\rm eff}(\bm \sigma)+J \SOMMA{a}{}\xi^1\sigma^{(a)}\Big)\right] \notag 
    \\
    &=\lim_{n\to 0}\dfrac{1}{n}\ln \left\{    \int \D u\, \int \left[\prod_{k=1}^{n/\theta}\D v_k\right] \sum_{\bm \sigma} \exp \left[ \frac{\b P (1+\rho)^{P/2-1} }{2}  \hat{\eta}_M \n^{P-1} \,\SOMMA{a=1}{n}\sigma^{(a)}+J \SOMMA{a=1}{n}\xi^1\sigma^{(a)} \,\right.\right. \notag 
    \\
        &\left.\left.+\b u\,\SOMMA{a=1}{n}\sigma^{(a)}\sqrt{\gamma \frac{(1+\sqrt{\rho_P})^2}{(1+\rho)^{P}} \frac{P}{2}\bar{q}_1^{P-1}}+\b \sum_{k=1}^{n/\theta}v_k\SOMMA{b=1}{\theta}\sigma^{((k-1)\theta+b)}\sqrt{\gamma \frac{(1+\sqrt{\rho_P})^2}{(1+\rho)^{P}} \frac{P}{2}(\bar{q}_2^{P-1}-\bar{q}_1^{P-1})}  \,\right]\right\} 
        \end{align}
where we have set $\D u = du \,e^{-u^2/2}/\sqrt{2\pi}$ and $\D v_k = dv_k \,e^{-v_k^2/2}/\sqrt{2\pi}$ for $k=1, \hdots, n/\theta$. \new{Now, writing 
$$\SOMMA{a=1}{n}\sigma^{(a)}=\SOMMA{k=1}{n/\theta}\SOMMA{b=1}{\theta}\sigma^{((k-1)\theta+b)},$$}
$$
$$
\new{the sum over the spin configuration can be done explicitly, finding
\begin{align}
    &=\lim_{n\to 0}\dfrac{1}{n}\ln \left\{    \int \D u\, \prod_{k=1}^{n/\theta}\int \D v_k\,\prod_{b=1}^{\theta}\sum_{\sigma^{((k-1)\theta+b)}} \exp \left( \frac{\b P (1+\rho)^{P/2-1} \hat{\eta}_M \n^{P-1}\sigma^{((k-1)\theta+b)}}{2}  \,\right.\right. \notag 
    \\
    &\hspace{1cm} +J \xi^1\sigma^{((k-1)\theta+b)}+\b u\,\sigma^{((k-1)\theta+b)}\sqrt{\gamma \frac{(1+\sqrt{\rho_P})^2}{(1+\rho)^{P}} \frac{P}{2}\bar{q}_1^{P-1}} \notag \\
        &\hspace{1cm}\left.\left.  +\b v_k\sigma^{((k-1)\theta+b)}\sqrt{\gamma \frac{(1+\sqrt{\rho_P})^2}{(1+\rho)^{P}} \frac{P}{2}(\bar{q}_2^{P-1}-\bar{q}_1^{P-1})}  \,\right)\right\} \notag \\
        &=\lim_{n\to 0}\dfrac{1}{n}\ln \left\{    
  \int \D u\, \prod_{k=1}^{n/\theta}\left[\int\D v_k 2^\theta
 \cosh^\theta\left( \frac{\b P (1+\rho)^{P/2-1}}{2}  \hat{\eta}_M \n^{P-1} +J \xi^1 \,\right.\right.\right.
 \notag 
    \\
        &\hspace{1cm}\left.\left.\left. +\b u\,\sqrt{\gamma \frac{(1+\sqrt{\rho_P})^2}{(1+\rho)^{P}} \frac{P}{2}\bar{q}_1^{P-1}}+\b v_k
        \sqrt{\gamma \frac{(1+\sqrt{\rho_P})^2}{(1+\rho)^{P}} \frac{P}{2}\left(\bar{q}_2^{P-1}-\bar{q}_1^{P-1}\right)}  \,\right)\right]\right\} \notag
        \end{align}}
Now applying the $n\to 0$ limit \cite{SK1975} and exploiting the relation $\lim\limits_{n\to 0}\dfrac{1}{n} \ln\mathbb{E}_{Y^{(1)}}[f^n(Y^{(1)})]=\mathbb{E}_{Y^{(1)}}[\ln f(Y^{(1)})]$, we get
\begin{align}
    &= \lim_{n\to 0}\dfrac{1}{n}\ln \left\{   \mathbb{E}_u\left[\mathbb{E}_v  \,2^\theta\cosh^\theta\left(\b u\,\sqrt{\gamma \frac{(1+\sqrt{\rho_P})^2}{(1+\rho)^{P}} \frac{P}{2}\bar{q}_1^{P-1}} \,\right.\right.\right. \notag 
    \\
        &\left.\left.\left.\textcolor{white}{+\mathbb{E}_{\bm \xi^1}} +\b v\sqrt{\gamma \frac{(1+\sqrt{\rho_P})^2}{(1+\rho)^{P}} \frac{P}{2}(\bar{q}_2^{P-1}-\bar{q}_1^{P-1})}  \,+ \frac{\b P (1+\rho)^{P/2-1} }{2}  \hat{\eta}_M \n^{P-1} +J\xi^1\,\right)\right]^{n/\theta}\right\} \notag 
        \\
        &= \dfrac{1}{\theta}\mathbb{E}_u\ln   \left\{\mathbb{E}_v  \,2^\theta\cosh^\theta\left[\b\left(u\,\sqrt{\gamma \frac{(1+\sqrt{\rho_P})^2}{(1+\rho)^{P}} \frac{P}{2}\bar{q}_1^{P-1}} +v\sqrt{\gamma \frac{(1+\sqrt{\rho_P})^2}{(1+\rho)^{P}} \frac{P}{2}(\bar{q}_2^{P-1}-\bar{q}_1^{P-1})}  \right.\right.\right.\notag
        \\
        &\,\left.\left.\left.+ \frac{ P (1+\rho)^{P/2-1}\hat{\eta}_M }{2}   \n^{P-1} +J\xi^1\,\right)\right]\right\} \notag 
\end{align}

Finally, inserting this expression in \eqref{eq:rt_free-ans} and denoting with $\mathbb{E}_{Y^{(1)}}[.]$ and $\mathbb{E}_{Y^{(2)}}[.]$ the Gaussian averages w.r.t. $Y^{(1)}$ and $Y^{(2)}$ respectively, we reach the following expression for the free energy 
\begin{equation}
\label{eq:rt_final}
\begin{array}{lll}
     -\b\mathcal{F}^{(P)}(J)
     =&\ln 2+ \dfrac{1}{\theta}\mathbb{E}_{\bm\xi^1}\mathbb{E}_{(\bm\eta^1|\bm\xi^1)}\mathbb{E}_{Y^{(1)}}\ln\mathbb{E}_{Y^{(2)}} \left\{ \cosh^\theta\left[J\xi^1+\beta '\dfrac{P}{2}\n^{P-1}(1+\rho)^{P/2-1} \etaM \right. \right.
        \\\\
     &\left. \left.+Y^{(1)} \beta '\sqrt{\gamma
     \dfrac{P}{2}\dfrac{(1+\sqrt{\rho_P})^2}{(1+\rho)^{P}} \q_1^{^{P-1}}} +Y^{(2)} \beta '\sqrt{\gamma
     \dfrac{P}{2}\dfrac{(1+\sqrt{\rho_P})^2}{(1+\rho)^{P}} ( \q_2^{^{P-1}} -  \q_1 ^{^{P-1}})}\right]\right\}\notag \\
     &-\dfrac{{\b}}{2}(P-1)(1+\rho)^{P/2}\n^P + \dfrac{{\b}^2}{4}\gamma\dfrac{(1+\sqrt{\rho_P})^2}{ (1+\rho)^{P}}\left[1-P\q_2^{P-1}+(P-1)\q_2^P\right]
         \\\\
    &-\dfrac{{\b}^2}{4}\gamma\dfrac{(1+\sqrt{\rho_P})^2}{ (1+\rho)^{P}}(P-1)\theta(\q_2^P-\q_1^P),
\end{array}
\end{equation}
where $\n,\q_1$ and $\q_2$ must fulfill the same self-consistency equations \eqref{eq:High_store_self_n_q} obtained by using Guerra's interpolation technique, as reported in Proposition \ref{P_quenched}.

\subsection{Results}
\label{sec:results-unsup}
%

\begin{figure}[t]
    \centering
\includegraphics[width=15.5cm]{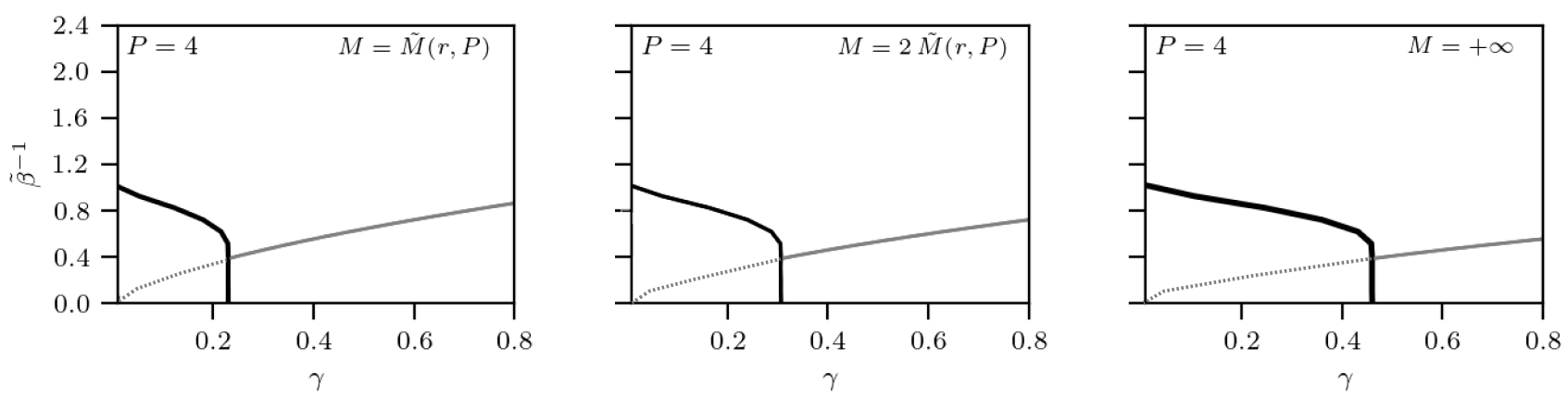}
    \caption{Phase diagram of the unsupervised DAM model with $P=4$, in the parameters space ($\gamma$, $\tilde{\beta}$), where $\gamma$ is the storage load defined in \eqref{eq:alphaPP-1} and $\tilde\beta=\b/(1+\rho)^{P/2}$ is the scaled inverse noise, where $\rho$ is the 
dataset entropy defined in \eqref{eq:rho}.
    Different panels refer to different sizes $M$ of the training set, as shown in the legend, where $ \tilde M(r,P) = r^{-2P}$ and $r$ has been set to $r=0.2$. For high values of the noise $\tilde{\beta}^{-1}$, the system is ergodic, while lowering the noise two phases appear: a retrieval phase at small $\gamma$ (below the black line), and a spin glass phase (below the grey line) at high values of $\gamma$. The spin-glass phase within the retrieval region (delimited by the dashed line) is unstable.  As $M$ increases, the spin glass region shrinks and the retrieval region expands.}
    \label{fig:unsup}
\end{figure}

\begin{figure}[t]
    \centering
 \includegraphics[width=15.5cm]{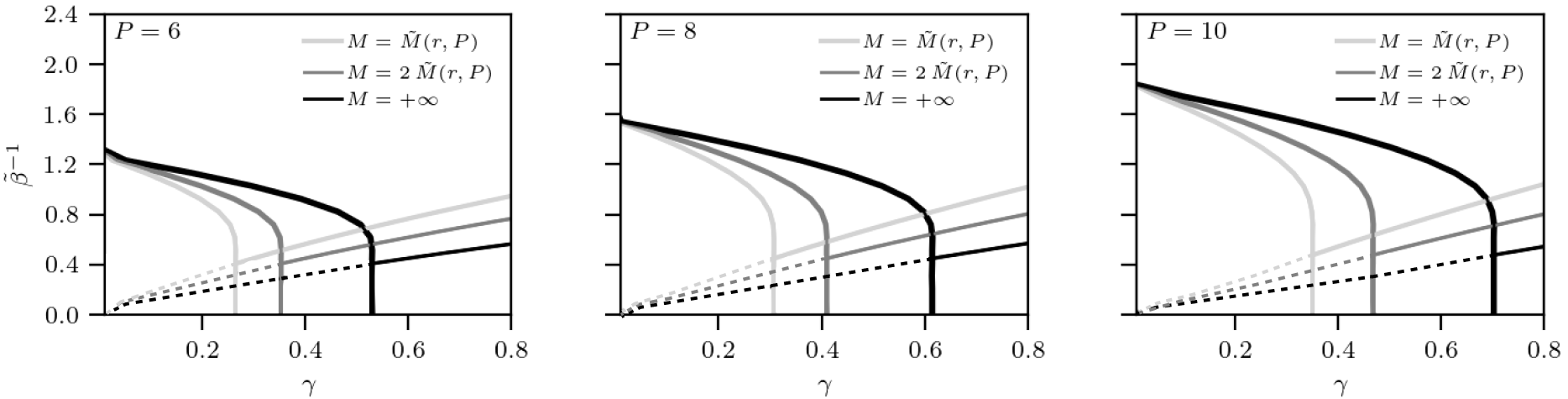}
    \caption{Phase diagrams 
    of the unsupervised DAMs model for $P=6, 8, 10$, and different values of $M$, as shown in the legends. $\tilde M(r,P)$ and $r$ are as in Figure \ref{fig:unsup}. As $P$ increases, the spin glass region shrinks and the retrieval region expands. }
    \label{fig:unsup_vari_P}
\end{figure}

\begin{figure}[t]
     \centering
     \begin{subfigure}[b]{0.48\textwidth}
         \centering
         \includegraphics[height=5cm,width=7.5cm]{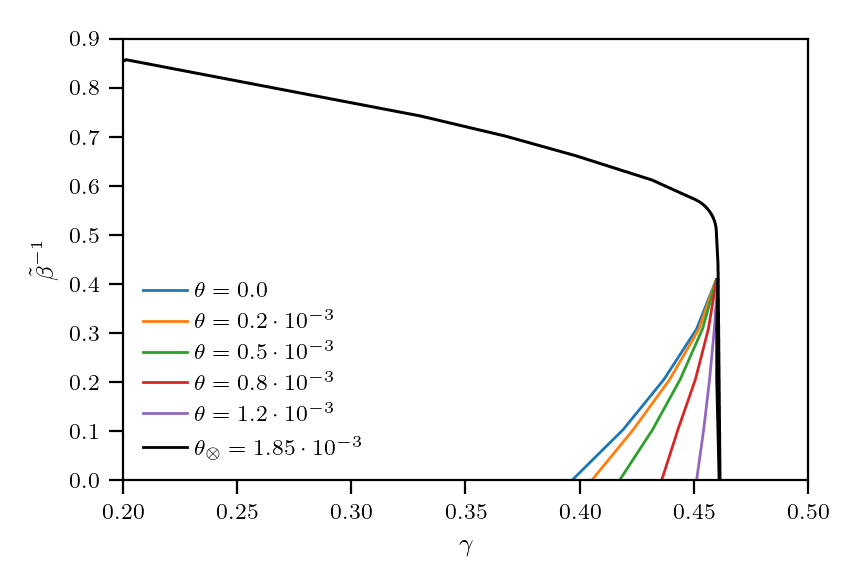}
     \end{subfigure}
     \begin{subfigure}[b]{0.48\textwidth}
         \centering
         \includegraphics[height=5cm,width=7.5cm]
         {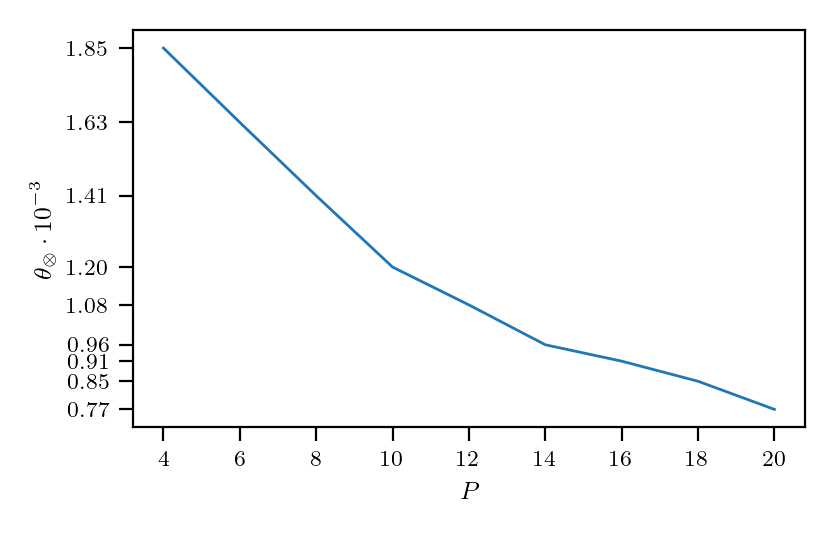}
     \end{subfigure}
        \caption{Left: Critical line separating the retrieval (left) from the spin-glass or ergodic region (right) for different values of $\theta$. For $\theta=0$, where the theory is RS, the retrieval region exhibits a re-entrance.As $\theta$ is increased the retrieval region expands, reaching its maximum at  $\theta=\theta_\otimes$, where the re-entrance disappears completely.
        Right: $\theta_\otimes$ versus $P$. As $P$ increases the value of $\theta_{\otimes}$ decreases, indicating that the larger $P$ the better the RS approximation is expected to be.}
        \label{fig:interpolation_RS_RSB}
\end{figure}

Solving numerically the self consistency equations \eqref{eq:High_store_self_n_q}, we obtain the phase diagram shown in Figure \ref{fig:unsup}, in the parameters space ($\gamma$, $\tilde{\beta}$), where $\gamma$ is the storage load defined in \eqref{eq:alphaPP-1} and $\tilde\beta=\b/(1+\rho)^{P/2}$ is a scaled inverse temperature, where $\rho$ is the 
dataset entropy defined in \eqref{eq:rho}. Panels show results for $P=4$ and three different values of $M$, as shown in the legend. 
The result for $M\to \infty$ follows from the analysis carried out in Sec \ref{sec:extensions_uns}.
In each panel, 
the grey curve marks the transition from the ergodic phase with $\bar{m}=\bar{q}_1=\bar{q}_2=0$
(above) to the 
spin glass phase, where either $\bar{q}_1$ or $\bar{q}_2$ becomes non-zero (below).   
The black curve marks the transition from the retrieval phase with $\bar{m}\neq 0$ (left) to $\bar{m}=0$ (right). 
The spin glass solution within the retrieval region is always unstable and it is delimited by the dotted curve (we refer to this as the instability region).
In Figure \ref{fig:unsup_vari_P} we show the results for $P=6, 8, 10$, as shown in the legend. 
As $P$ and $M$ grow, the spin glass region shrinks and the retrieval region expands. 
For $M=+\infty$ the critical storage lines become equal to those of traditional DAM models where archetypes are encoded in the interactions, rather than their noisy examples \cite{Albanese2021}. 
This is as expected: when an infinite 
number of examples is provided, the system reaches 
the same performance as a neural network 
where archetypes are stored in the interactions directly. For finite $M$, however, the network's ability to retrieve the archetypes degrades at values of the storage load which are lower than the storage capacity in traditional DAM models.
Both Figures \ref{fig:unsup} and \ref{fig:unsup_vari_P} 
have been obtained for the value of $\theta=\theta_\otimes$ which maximizes the retrieval region, as shown in Figure \ref{fig:interpolation_RS_RSB} (left panel). In the latter,  
we show the line that separates the retrieval region from the spin-glass or the ergodic region, for different values of $\theta$. 
For $\theta=0$ (which corresponds to the RS theory), the line exhibits a re-entrance,
as commonly observed in spin-glasses and associative memories \cite{SK1975,AGS-1987,Crisanti-RSB,Albanese2021}. As $\theta$ is increased, the retrieval region expands, reaching its maximum at 
$\theta=\theta_\otimes$, where the re-entrance disappears completely, showing the same qualitative behaviour as in the Hopfield model \cite{AGS} and traditional dense associative memories \cite{Albanese2021}. Increasing $\theta$ above $\theta_\otimes$ the re-entrance appears again and gets more pronounced as $\theta$ approaches $1$, where the transition line becomes identical to the RS (and the $\theta=0$) case,  as expected. 
In Figure \ref{fig:interpolation_RS_RSB} (right panel) we show the dependence of $\theta_\otimes$ on $P$.
As $P$ increases, the value of $\theta_{\otimes}$ decreases. This is as expected from spin-glass theory, as for large $P$, $P$-spin models are known to converge to the REM model \cite{derrida1981random}, which is RS. 
Interestingly, we find that $\theta_\otimes$ takes the same value as in the DAM model considered in \cite{Albanese2021}, for all values of $M$, hence the optimal $1$RSB breaking parameter 
is not influenced by the use of corrupted examples rather than archetypes.  

\subsubsection{Limiting cases: $M\to \infty$ and $\beta\to \infty$}
\label{sec:extensions_uns}
Finally, we consider two instructive limiting cases. One is the limit $M\to\infty$,  where the number of available examples is large. Although idealised, this scenario is becoming less utopian nowadays in a number of applications, and it is instructive as an explicit relation between the archetype magnetization and the mean magnetization of the examples naturally emerges. The second scenario is the zero noise limit $\beta\to\infty$, where the information processing capabilities of the network are expected to be maximal. 
We now state the following 
\begin{corollary}
\label{cor:large}
In the limit $M\to\infty$, the 1RSB self consistent equations for the order parameters of the unsupervised DAM model are
\begin{align}
    \n&= \dfrac{\m}{1+\rho} +\beta ' \dfrac{P}{2}\rho(1+\rho)^{P/2-2}\left[1+(\theta-1)\q_2-\theta \q_1\right]\n^{^{P-1}},\notag
    \\
    \q_1&= \mathbb{E}_{Y^{(1)}} \left[\dfrac{\mathbb{E}_{Y^{(2)}} \cosh^\theta g(\beta, \gamma, \bm Y) \tanh g(\beta, \gamma, \bm Y)}{\mathbb{E}_{Y^{(2)}} \cosh^\theta g(\beta, \gamma, \bm Y)} \right]^2, \notag
         \\
    \q_2 &= \mathbb{E}_{Y^{(1)}} \left[\dfrac{\mathbb{E}_{Y^{(2)}} \cosh^\theta g(\beta, \gamma, \bm Y) \tanh^2 g(\beta, \gamma, \bm Y)}{\mathbb{E}_{Y^{(2)}} \cosh^\theta g(\beta, \gamma, \bm Y)} \right],
    \label{eq:selfLarge}
    \\
    \m&= \mathbb{E}_{Y^{(1)}}\left[\dfrac{\mathbb{E}_{Y^{(2)}} \cosh^\theta g(\beta, \gamma, \bm Y) \tanh g(\beta, \gamma, \bm Y)}{\mathbb{E}_{Y^{(2)}} \cosh^\theta g(\beta, \gamma, \bm Y)}\right].\notag
\end{align} 
where 
 \begin{align}
    g(\beta, \gamma, \bm Y)=&\beta '\dfrac{P}{2}\n^{^{P-1}}(1+\rho)^{P/2-1} + Y^{(2)} \b \sqrt{
             \gamma\dfrac{ P}{2} \dfrac{(1+\sqrt{\rho_P})^2}{(1+\rho)^P}(\q_2^{^{P-1}} - \q_1 ^{^{P-1}})} \notag \\
             &+ \beta 'Y^{(1)}\sqrt{\rho\left(\dfrac{P}{2}\n^{^{P-1}}(1+\rho)^{P/2-1}\right)^2+  \gamma
    \dfrac{P}{2}\dfrac{(1+\sqrt{\rho_P})^2}{(1+\rho)^{P}}\,\q_1^{^{P-1}}\;}
    \label{eq:g_of_unsuper_n_app}
 \end{align}
 \normalsize
 and $\b=2\beta /P!$.
\end{corollary}

\begin{proof}
In the limit $M\to \infty$, we can apply the CLT to the sum of the examples defined in \eqref{eq:etaM}, to write 
\begin{equation}
    \etaM\sim \xi^1(1+\lambda\sqrt{\rho}),
    \label{eq:CLT_unsup}
\end{equation}
where $\lambda$ is a standard Gaussian variable $\lambda\sim \mathcal{N}(0,1)$. 

%
Inserting \eqref{eq:CLT_unsup} in \eqref{eq:g} 
and in the expression for $\bar{n}$ provided in \eqref{eq:High_store_self_n_q}, 
we get 
\begin{align}
    g(\beta, \gamma, \bm Y, \lambda, \xi^1) =& \beta '\dfrac{P}{2}\n^{P-1}(1+\rho)^{P/2-1}\xi^1(1+\sqrt{\rho} \lambda)+Y^{(1)} \beta '\sqrt{\dfrac{\gamma P (1+\sqrt{\rho_P})^2}{2(1+\rho)^{P}}  \q_1^{^{P-1}}} \notag \\
    &+Y^{(2)} \beta '\sqrt{
    \dfrac{\gamma (1+\sqrt{\rho_P})^2 P}{2(1+\rho)^{P}} ( \q_2^{^{P-1}} -  \q_1 ^{^{P-1}})}\,.
\end{align}
and
$$\n = \frac{1}{(1+\rho)} \mathbb{E}_{\xi^1}\mathbb{E}_{\lambda}\mathbb{E}_{Y^{(1)}} \left[\dfrac{\mathbb{E}_{Y^{(2)}}\cosh^{\theta} g(\beta,\gamma, \bm Y,\lambda, \xi^1) \tanh^{\theta} g(\beta,\gamma, \bm Y,\lambda,\xi^1) }{\mathbb{E}_{Y^{(2)}}\cosh^{\theta} g(\beta,\gamma, \bm Y,\lambda,\xi^1)} \cdot {\xi^1(1+\lambda \sqrt{\rho})}\right]$$
Hence, by applying to the standard Gaussian variable $\lambda$ the Stein's lemma, which states that for a standard Gaussian variable $J\sim N(0, 1)$ and a generic function $f(J)$, for which the two expectations $\mathbb{E}\left( J f(J)\right)$ and $\mathbb{E}\left( \partial_J f(J)\right)$ both exist, one has 
\begin{equation}
\label{eqn:gaussianrelation2Pspin}
\mathbb{E} \left[ J f(J)\right]= \mathbb{E} \left[ \frac{\partial f(J)}{\partial J}\right],
\end{equation}
and explicitly averaging over $\xi$ we get the self-consistency equations in \eqref{eq:selfLarge}.

\end{proof}

\new{Finally, it has been shown in
\cite{super}, within a RS analysis, that neglecting the second term in the  equation for $\bar{n}$, given in \eqref{eq:selfLarge}, has negligible impact on the solutions of the self-consistency equations, in the relevant regime of low noise, where the network works as an associative memory. As the equation for $\bar{n}$ is the same in the RS theory and in the $1$RSB theory that we are considering here, the 
same truncation of $\bar{n}$ can be adopted here}
\begin{equation}
    \n = \dfrac{\m}{1+\rho}
    \label{eq:trunc_n}
\end{equation}
This leads to a simplified expression for $g$ that is
 \begin{align}
    g(\beta, \gamma, \bm Y)=&\tilde\beta\dfrac{P}{2}\m^{^{P-1}}(1+\rho)^{P/2-1} + \tilde\beta Y^{(2)}  \sqrt{
             \gamma\dfrac{ P}{2} (1+\sqrt{\rho_P})^2(\q_2^{^{P-1}} - \q_1 ^{^{P-1}})} \notag \\
             &+ \tilde\beta Y^{(1)}\sqrt{\rho\left(\dfrac{P}{2}\m^{^{P-1}}\right)^2+  \gamma
    \dfrac{P}{2}(1+\sqrt{\rho_P})^2\,\q_1^{^{P-1}}\;}\;.
    \label{eq:g_of_unsuper_n_app_unsup_troncata}
 \end{align}
 where we re-scaled the noise $\tilde\beta=\b/(1+\rho)^{P/2}$.
\new{
Solving numerically the self consistency equations \eqref{eq:selfLarge} where the equation for $\bar{n}$ is replaced with its truncated version 
\eqref{eq:trunc_n}, leads to the phase diagram 
shown by the lines $M=+\infty$ in Figs. \ref{fig:unsup} and \ref{fig:unsup_vari_P}.}
\\[5mm]
Next, we turn to the analysis of the ground state, i.e. to the limit $\beta \to \infty$, and we state the following 

\begin{corollary}
\label{cor:nulltemp_rho}
In the limit $\beta\to\infty$ and for $M\gg 1$, the $1$RSB self consistency equations for the order parameters of the unsupervised DAM model are 
\begin{align}
    \m &= 1- 2 \mathbb{E}_{Y^{(1)}} \left[ \dfrac{1}{1+\exp\left( 2D(A_1 + A_2 Y^{(1)})\right)\mathcal{Q}(A)}\right] \label{eq:m_beta_rho}\\
    \Delta \q&=\q_2-\q_1 = 4 \mathbb{E}_{Y^{(1)}} \left\{ \dfrac{\exp\left( 2D(A_1 + A_2 Y^{(1)})\right)\mathcal{Q}(A)}{\left[1+\exp\left( 2D(A_1 + A_2 Y^{(1)})\right)\mathcal{Q}(A)\right]^2}\right\} \label{eq:q_beta_rho}
\end{align}
where $D=\tilde\beta \theta$ and
\begin{equation}
    \begin{array}{lll}
    &\mathcal{Q}(A)= \dfrac{1+\mathrm{erf}(K^{+})}{1+\mathrm{erf}(K^{-})}, \ \ \ &K^{\pm}= \dfrac{D A_3^2 \pm (A_1 + A_2 Y^{(1)})}{A_3 \sqrt{2}}, 
    \\\\
    &A_1 = \dfrac{P}{2}\m^{P-1}, \ \ \ & A_2 = \sqrt{\rho A_1^2+\gamma(1+\sqrt{\rho_P})^2\dfrac{P}{2}},
    \\\\
    &A_3 = \sqrt{\gamma(1+\sqrt{\rho_P})^2\dfrac{P(P-1)}{2}\Delta\q}.
    \end{array}
\end{equation}

\end{corollary}

We report the proof of this Corollary in Appendix \ref{app:proofnulltemp}. 

\begin{figure}
    \centering
    \includegraphics[width=15cm]{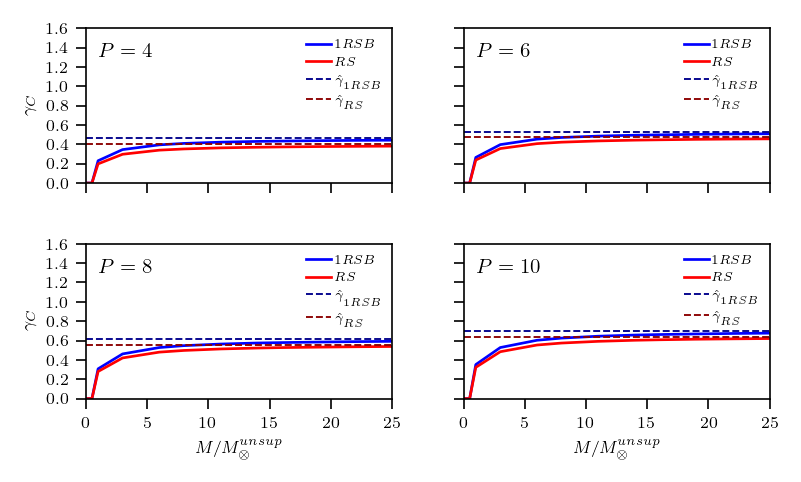}
    \caption{Critical storage capacity $\gamma_c$ in the ground state for the unsupervised DAM model, for example noise $r=0.2$, as a function of the ratio $M/M_\otimes^{unsup}$ both for the RS (red line) and $1$RSB (blue line) assumptions. We notice that $\gamma_c$ increases with $M$ and, for $M \gg M_{\otimes}^{unsup}$, it saturates to the critical load ($\hat\gamma$) found numerically for the RS case in \cite{EmergencySN}.}
    \label{fig:null_unsup}
\end{figure}

Next we compute, by numerically solving \eqref{eq:m_beta_rho}-\eqref{eq:q_beta_rho}
, the ground-state critical storage capacity $\gamma_{C}$ 
beyond which a black-out scenario emerges, namely $\m\neq 0$ for $\gamma<\gamma_C$ and $\m= 0$ for $\gamma>\gamma_C$. 
In Fig. \ref{fig:null_unsup} we plot $\gamma_C$ as a function of the ratio between the number $M$ of experienced examples and the minimum number $M_\otimes$ of examples required by the unsupervised DAM model (with $P>2$) to correctly learn and retrieve an archetype, within the RS theory, which has been proved to be    
\begin{align}
     M^{unsup}_{\otimes}(r, P, \gamma)= \gamma \dfrac{1}{P} \dfrac{1}{r^{2P}}.
\end{align}
 in \cite{unsup}. 
In order to ascertain whether replica symmetry breaking alters such threshold, we plot $\gamma_C$
within both the $1$RSB assumption (blue line) and the RS theory (red line). We see that $\gamma_C$ becomes non-zero (meaning that retrieval can occur) at the same value of $M=M_\otimes$ for the RS and the $1$RSB theory. This shows that the minimum number 
of examples that a network needs to accomplish retrieval of the archetype is the same within the RS or the $1$RSB assumption. 

Results show that, as in the standard Hebbian storage \cite{AABO-JPA2020,Crisanti-RSB,Steffan-RSB}, the phenomenon of replica symmetry breaking induces a mild improvement in terms of the maximal storage.


\section{`Supervised' dense associative memories}
\label{sec:sup}
In this section we analyse the information processing capabilities of the DAM model in the so-called `supervised' setting, where the dataset given to the network is now split into different categories, one for each archetype $\bm{\xi}^\mu$, with $\mu=1\ldots P$.
Again, we 
analyze the model, defined in Sec. \ref{sec:def-sup}, via both Guerra's interpolation techniques (Sec. \ref{GuerraSup}) and Parisi's replica approach (Sec. \ref{ParisiSup}), at the first step of replica-symmetry breaking. 
In Appendix \ref{Sec:labile}
we derive the instability line of the RS theory.

\subsection{Model and definitions}
\label{sec:def-sup}
As in the unsupervised DAM model, we consider a network of $N$ Ising neurons $\si \in \{ -1, +1\}$, $i=1, \hdots , N$, interacting via $P$-node interactions. We assume to have $K$ Rademacher archetypes $\bm \xi^\mu \in \{-1, +1\}^N$, with $\mu=1, \hdots, K$, defined as $N$-dimensional vectors with entries drawn randomly and independently from the distribution \eqref{eq:xi}. In addition, we assume to have $M$ examples $\eta^{\mu,a} \in \{-1, +1\}^N$ for each archetype, which are corrupted version of the archetypes, with entries distributed according to  \eqref{eq:Bernoulli}. We shall refer to this model as the supervised DAM model. 

\begin{definition}   
\label{def:H_sup}
The Hamiltonian of the supervised DAM model is
\begin{align}
    \mathcal H^{(P)}_{N}(\boldsymbol{\sigma} \vert \bm \eta)=& -\dfrac{1}{ \mathcal{R}^{P/2}N^{P-1}M^P}\SOMMA{\mu=1}{K}\SOMMA{a_1,\cdots,a_P=1}{M,\cdots,M}\left(\SOMMA{i_1<\cdots<i_P}{N,\cdots,N}\eta^{\mu, a_1}_{i_1}\cdots\eta^{\mu, a_P}_{i_P}\sigma_{i_1}\cdots\sigma_{i_P}\right) 
    \label{eq:H_Psup}
\end{align}
where the constant $\mathcal{R}:=r^2 + \frac{1-r^2}{M}$ in the denominator of the r.h.s. is included for mathematical convenience and the factor $N^{P-1}$ ensures the Hamiltonian to be ${\mathcal O}(N)$, as explained previously for the unsupervised model, see Eq. \eqref{def:H_PHopEx}.
\end{definition}
We highlight the hidden role of a teacher that, before providing the dataset to the network, has grouped examples pertaining to the same archetype together (hence the proliferation of summations in the cost function \eqref{eq:H_Psup} with respect to \eqref{def:H_PHopEx}).

As before, we will focus (without loss of generality)
on the ability of the network to store and retrieve the first archetype $\boldsymbol \xi^1$, hence the Mattis magnetization $m_1(\bm{\sigma})$ provided in \eqref{eq:def_m} remains a relevant order parameter. However, the set of order parameters for the examples, previously given by $n_{1,a}(\bold{\sigma})$ with $a=1\ldots M$ (see \eqref{eq:def_n}) have now to be substituted with a single order parameter
\begin{equation}
n_1(\bm\sigma)=\dfrac{r}{\mathcal{R}}\dfrac{1}{N M}\SOMMA{i,a=1}{N,M}\eta_i^{1,a}\sigma_i. 
\end{equation}
Its probability distribution, in the thermodynamic limit, is assumed self-averaging, namely
\begin{equation}
\lim_{N \rightarrow + \infty} \mathbb{P}_N(n_1|\bm\sigma) = \delta (n_1(\bm\sigma) - \bar{n}),
\label{limfornPspin_sup}
\end{equation}
as for the Mattis magnetization, while the distribution for the overlap defined in \eqref{eq:def_q} is still assumed bimodal as in Assumption \eqref{def:HM_RSBPspin} (see \eqref{limforq2Pspin}).
As for the unsupervised model, we add an extra term in the cost function, namely $J \sum_i \xi_i^1 \si$, in order to generate the moments of the Mattis magnetization $m_1$ by taking the derivatives of the quenched free energy w.r.t. $J$ and, as this term is not part of the original Hamiltonian, it  will be set to zero at the end of the calculations. 
Therefore, we write the partition function as
\begin{align}
    \mathcal{Z}^{(P)}_{N} (\bm \eta)
    &= \lim_{J \rightarrow 0} \mathcal{Z}^{(P)}_{N,K,M,\beta}( \bm \eta ;J) \notag \\
    &= \lim_{J \rightarrow 0} \sum_{\bm \sigma}  \exp \left\{ J \sum_i \xi_i^1 \si +\beta'\dfrac{N}{2}(1+\rho)^{P/2}n_1^{P}(\boldsymbol{\sigma})\notag \right.\\
    &\left.+\dfrac{\beta'P!}{2N^{P-1}\mathcal{R}^{P/2} }\SOMMA{\mu>1}{K}\SOMMA{i_1<\cdots <i_{P}}{N,\cdots,N}\left[ \left(\dfrac{1}{M}\SOMMA{a_1=1}{M}\eta_{i_1}^{\mu,a_1} \right)\hdots \left(\dfrac{1}{M}\SOMMA{a_P=1}{M}\eta_{i_{P}}^{\mu,a_{P}}\right) \right]\sigma_{i_1}\cdots\sigma_{i_{P}} \right\}
    \label{eq:partition}
\end{align}
where $\rho$ is the dataset entropy defined in \eqref{eq:rho}, $\beta':=2\beta/P!$ and for the first pattern, $\mu=1$, we have used the relation $P!\sum_{i_1 <\cdots < i_P}=\sum_{i_1, \cdots, i_P}$ and neglected terms which vanish in the thermodynamic limit. 
Next, we apply the CLT to the variables in the round brackets in \eqref{eq:partition}, and rewrite each term as 
\begin{equation}
    \dfrac{1}{M}\SOMMA{a=1}{M}\eta^{\mu,a}_i= r\xi_i^\mu \Big(1+\phi_i^\mu \sqrt{\rho}\Big)\;\;\;\;\mathrm{with}\;\;\;\;\phi_i^\mu\sim \mathcal{N}(0,1)\,.
\end{equation}
Replaicing the previous in the expression inside the square brackets in \eqref{eq:partition}
we obtain
$$
\SOMMA{\mu>1}{K}\left[ \left(\dfrac{1}{M}\SOMMA{a_1=1}{M}\eta_{i_1}^{\mu,a_1} \right)\hdots \left(\dfrac{1}{M}\SOMMA{a_P=1}{M}\eta_{i_{P}}^{\mu,a_{P}}\right) \right]=
r^P\SOMMA{\mu>1}{K}\xi_{i_1}^\mu\ldots \xi_{i_P}^\mu \left(1+\phi_{i_1}^\mu\sqrt{\rho}\right)\ldots
\left(1+\phi_{i_P}^\mu\sqrt{\rho}\right)
$$
and apply again the CLT now over the $\mu$-summation, we can write 
$$
\frac{1}{K}\SOMMA{\mu>1}{K}\left[ \left(\dfrac{1}{M}\SOMMA{a_1=1}{M}\eta_{i_1}^{\mu,a_1} \right)\hdots \left(\dfrac{1}{M}\SOMMA{a_P=1}{M}\eta_{i_{P}}^{\mu,a_{P}}\right) \right]= \lambda_{i_1,\hdots,i_P} \sqrt{\dfrac{r^{2P}(1+\rho)^{P}}{K}}
$$
where $\lambda_{i_1,\hdots,i_P}\sim \mathcal{N}(0,1)$.\\
Inserting all back in \eqref{eq:partition} we get
\begin{align}
    \mathcal{Z}^{(P)}_{N,K,M,\beta} (\bm \eta^1,\bm\lambda;J)
    &=  \exp \left\{ J \sum_i \xi_i^1 \si +\beta'\dfrac{N}{2}(1+\rho)^{P/2}n_1^{P}(\boldsymbol{\sigma})+\dfrac{\beta'P!\sqrt{K}}{2N^{P-1} }\SOMMA{i_1<\cdots< i_{P}}{N,\cdots,N} \lambda_{i_1,\hdots,i_{P}} \sigma_{i_1}\cdots\sigma_{i_{P}} \right\}\,.
    \label{eq:partition_supervised}
\end{align}

In the next subsection we calculate the free energy of the system in the thermodynamic limit using Guerra’s interpolation technique, assuming one step of replica symmetry breaking (1RSB).

\subsection{1RSB analysis via Guerra's interpolation technique}\label{GuerraSup}

As for the unsupervised case, the plan is to construct an interpolation between the original model and a simpler one-body model, whose statistical features are as close as possible to the original one, then solve the one body model and finally obtain the solution of the original model via the fundamental theorem of calculus.
Thus we define the following interpolating partition function

\begin{definition}
\label{def:part_Interpolante_sup}
Given the interpolating parameter $t \in [0,1]$, $A_1,\ A_2, \ \psi \in \mathbb{R}$ constants to be set a posteriori, and the i.i.d. standard Gaussian variables $Y_i^{(b)} \sim \mathcal{N}(0,1)$ for $i=1, \hdots , N$ and $b=1,2$, the Guerra's 1-RSB interpolating partition function for the DAM model, trained by a teacher, is given by 
\begin{equation}
\begin{array}{lll}
    \mathcal{Z}^{(P)}_2(\bm\eta^1,\bm \lambda,\bm Y; J, t)&\coloneqq \sommaSigma{\boldsymbol \sigma} \mathcal{B}_2^{(P)}(\bm \sigma \vert \bm \eta^1,\bm\lambda,\bm Y; J, t)
    \\\\
    &= \sommaSigma{\boldsymbol \sigma} \exp{\Bigg[}J \sum_{i=1}^N \xi_i^1 \si+\dfrac{t \beta 'N}{2M}\left(1+\rho\right)^{P/2}n_{1}^{^P}(\bm\sigma) +\psi(1-t)Nn_{1}(\bm\sigma)
         \\\\
    &+\sqrt{t}\dfrac{\beta '\sqrt{K}}{N^{P-1}}\SOMMA{{i_1 <\cdots < i_P}}{N,\cdots,N}\lambda_{i_1,\hdots,i_P}\sigma_{_{i_1}}\cdots\sigma_{_{i_{_{P}}}}+\sqrt{1-t}\SOMMA{b=1}{2}A_b\SOMMA{i=1}{N}Y_i^{(b)}\sigma_i\Bigg]  .
     \label{def:partfunct_GuerraRS_sup}
\end{array}
\end{equation}
where $\mathcal{B}_2^{(P)}(\bm \sigma \vert \bm \eta^1,\bm\lambda,\bm Y; J, t)$ is denoted as Boltzmann factor. 
\end{definition}
As before, the introduction of the interpolating partition function gives rise to a generalized  measure, average and interpolating quenched free energy (that we do not repeat here). All these generalizations retrieve the standard definitions  when evaluated at $t=1$.
Following Guerra's method \cite{guerra_broken},
we must average out the fields ${\bf Y}^{(1)}$ and ${\bf Y}^{(2)}$ recursively, in the interpolating free energy resulting from the interpolating partition function \eqref{def:partfunct_GuerraRS_sup}, as already done in the previous Section for the unsupervised case, see Eqs. \eqref{eqn:Z1Pspin} \eqref{eqn:Z0_1RSBPspin} and \eqref{dimenticanza}.
Proceeding in the same way and omitting obvious 
details due to the similarity of the proofs in the supervised and unsupervised cases, we state directly the next 

\begin{proposition}\label{P_quenched_sup}
In the thermodynamic limit $N\to\infty$, within the $1$RSB assumption and
under the Assumption \ref{eq:seconda}, the quenched free energy for the supervised DAM model, reads as
\begin{equation}
\label{eq:pressure_Guerra_sup}
\begin{array}{lll}
    -\b\mathcal{F}^{(P)}     (J)
    &=\ln 2+ \dfrac{1}{\theta}\mathbb{E}_{\bm\xi^1}\mathbb{E}_{(\bm\eta^1|\bm\xi^1)}\mathbb{E}_{Y^{(1)}}  \ln\mathbb{E}_{Y^{(2)}}\left\{\cosh^\theta\left[J\xi^1+\beta '\dfrac{P}{2}\n^{P-1}(1+\rho)^{P/2-1} \etaM \right. \right.\\\\
    &\left. \left. +Y^{(1)} \beta '\sqrt{\gamma \dfrac{P}{2} \q_1^{^{P-1}}}+Y^{(2)} \beta '\sqrt{\gamma \dfrac{P}{2} \left(\q_2^{^{P-1}} -\q_1 ^{^{P-1}}\right)}\right]\right\} +\dfrac{{\b}^2}{4}\gamma\left[1-P\q_2^{P-1}+(P-1)\q_2^P\right]
         \\\\
    &-\dfrac{{\b}}{2}(P-1)(1+\rho)^{P/2}\n^P-\dfrac{{\b}^2}{4}\gamma(P-1)\theta(\q_2^P-\q_1^P),
\end{array}
\end{equation}
with  $\bar n$, $\q_1, \ \q_2$ fulfilling the following self-consistency equations
\begin{equation}
    \begin{array}{lll}
         \n=\dfrac{1}{1+\rho}\mathbb{E}_{\bm\xi^1}\mathbb{E}_{(\bm\eta^1|\bm\xi^1)} \mathbb{E}_{Y^{(1)}} \left[\dfrac{\mathbb{E}_{Y^{(2)}} \cosh^\theta g(\beta, \gamma, \bm Y, \bm\eta^1) \tanh g(\beta, \gamma, \bm Y, \bm\eta^1)}{\mathbb{E}_{Y^{(2)}} \cosh^\theta g(\beta, \gamma, \bm Y, \bm\eta^1)}\dfrac{1}{r} \chiM \right],
         \\\\
         \q_1= \mathbb{E}_{\bm\xi^1}\mathbb{E}_{(\bm\eta^1|\bm\xi^1)}\mathbb{E}_{Y^{(1)}} \left[\dfrac{\mathbb{E}_{Y^{(2)}} \cosh^\theta g(\beta, \gamma, \bm Y, \bm\eta^1) \tanh g(\beta, \gamma, \bm Y, \bm\eta^1)}{\mathbb{E}_{Y^{(2)}} \cosh^\theta g(\beta, \gamma, \bm Y, \bm\eta^1)} \right]^2,
         \\\\
         \q_2 = \mathbb{E}_{\bm\xi^1}\mathbb{E}_{(\bm\eta^1|\bm\xi^1)} \mathbb{E}_{Y^{(1)}} \left[\dfrac{\mathbb{E}_{Y^{(2)}} \cosh^\theta g(\beta, \gamma, \bm Y, \bm\eta^1) \tanh^2 g(\beta, \gamma, \bm Y, \bm\eta^1)}{\mathbb{E}_{Y^{(2)}} \cosh^\theta g(\beta, \gamma, \bm Y, \bm\eta^1)} \right],
    \end{array}
    \label{eq:High_store_self_n_q_sup}
\end{equation}
where 
\begin{align}
    g(\beta, \gamma, \bm Y, \bm\eta^1) =& \beta '\dfrac{P}{2}\n^{P-1}(1+\rho)^{P/2-1} \etaM+Y^{(1)} \beta '\sqrt{\gamma \dfrac{P}{2}  \q_1^{^{P-1}}}+Y^{(2)} \beta '\sqrt{\gamma \dfrac{P}{2} ( \q_2^{^{P-1}} -  \q_1 ^{^{P-1}})} 
    \label{eq:g_sup}
\end{align}
\normalsize
and $\b=2\beta /P!$.
Furthermore, as $\bar{m}=-\b \nabla_J \mathcal{F}^{(P)}(J)|_{J=0}$, we have
\begin{align}
\label{eq:High_store_self_m_sup}
    \m= \mathbb{E}_{\bm\xi^1}\mathbb{E}_{(\bm\eta^1|\bm\xi^1)}\mathbb{E}_{Y^{(1)}} \left[\dfrac{\mathbb{E}_{Y^{(2)}} \cosh^\theta g(\beta, \gamma, \bm Y, \bm\eta^1) \tanh g(\beta, \gamma, \bm Y, \bm\eta^1)}{\mathbb{E}_{Y^{(2)}} \cosh^\theta g(\beta, \gamma, \bm Y, \bm\eta^1)}\xi^1\right] .
    \end{align}
\end{proposition}
The proof of the aforementioned proposition is lengthy but the steps to follow are identical to the ones provided for the unsupervised case.

\subsection{1RSB analysis via Parisi's replica trick} 
\label{ParisiSup}
As stated in the previous Section, the core of this approach consists in writing the logarithm of the partition function as $\lim\limits_{n\to 0} (\mathbb{E}\mathcal{Z}(J)^n-1)/n$
where 
\begin{align}
     \mathbb{E}\mathcal{Z}_{N}^n(J) =& \mathbb{E} \sum_{\bm{\sigma}^1 \ldots \bm{\sigma}^n} \exp\left[\dfrac{\b P!}{2 \mathcal{R}^{P/2}N^{P-1}M^P}\SOMMA{a=1}{n}\SOMMA{\mu=1}{K}\left(\SOMMA{i_1 <\cdots < i_P}{N,\hdots,N}\SOMMA{A_1,\hdots,A_P}{M,\hdots,M}\eta^{\mu, A_1}_{i_1}\cdots\eta^{\mu, A_P}_{i_P}\sigma_{i_1}^{(a)}\cdots\sigma_{i_P}^{(a)}\right) \right.
     \notag
     \\
&\left.     +J\SOMMA{a,i}{n,N}\xi_i^1\sigma_i^{(a)}\right].
\end{align}
Proceeding as in the unsupervised case, we compute separately the signal term and the noise term. The former, accounting for the examples pertaining to the first archetype,  reads as
\begin{align}
     \mathbb{E}\mathcal{Z}_{signal}^n(J) =& \mathbb{E}_{\bm\xi^1}\mathbb{E}_{(\bm\eta^1|\bm\xi^1)} \sum_{\bm{\sigma}^1 \ldots \bm{\sigma}^n} \exp\left[\dfrac{\b N (1+\rho)^{P/2}}{2 }\SOMMA{a=1}{n}\left(\dfrac{r}{N M \R}\SOMMA{i}{N}\SOMMA{A}{M}\eta^{1, A}_{i}\sigma_{i}^{(a)}\right)^P
     \right.
     \notag
     \\
&\left.     +J\SOMMA{a=1}{n}\SOMMA{i=1}{N}\xi_i^1\sigma_i^{(a)}\right],
\end{align}
while the latter accounting for the examples  pertaining to all the other archetypes but the first one, can be written as

\begin{align}
     \mathbb{E}\mathcal{Z}_{noise}^n &= \mathbb{E}_{\bm\lambda} \sum_{\bm{\sigma}^1 \ldots \bm{\sigma}^n} \exp\left[\dfrac{\b P!\sqrt{K}}{2 N^{P-1}}\SOMMA{a=1}{n}\left(\SOMMA{i_1 <\cdots < i_P}{N,\hdots,N}\lambda_{i_1,\hdots,i_P}\sigma_{i_1}^{(a)}\cdots\sigma_{i_P}^{(a)}\right) \right]
\end{align}
where now $\mathbb{E}_{\bm \lambda}$ is the Gaussian average w.r.t. $\bm \lambda$. Performing the integration over $\bm\lambda$ and 
inserting the definitions of the order parameters as done for the unsupervised setting, we can write
\begin{align}
    &\mathbb{E}\mathcal{Z}_N^n(J) =  \int \prod_{a} d n_{1}^{(a)} \prod_{a} \dfrac{dz_{1}^{(a)}}{2\pi} \int \prod_{a,b} dq_{ab} \int \prod_{a,b} \dfrac{dp_{ab}}{2\pi} \exp\left( -N A[\bm Q, \bm P,\bm N, \bm Z] \right) 
\end{align}
where 
\begin{align}
    A[\bm Q,\bm P,\bm N,\bm Z]&= - \dfrac{i}{N} \sum_{a,b} p_{ab}q_{ab} -\dfrac{1}{4}{\b}^2\gamma \SOMMA{a,b}{n,n}q_{ab}^P - \dfrac{\b (1+\rho)^{P/2}}{2} \sum_{a} (n_{1}^{(a)})^P-\dfrac{i}{N}\sum_{a} n_{1}^{(a)} z_{1}^{(a)} \notag \\
    & 
    -\mathbb{E} \log \left[ \sum_{\bm \sigma} \exp\left( -\dfrac{i}{N} \sum_{a,b} p_{ab} \sigma^{(a)}\sigma^{(b)} - \dfrac{i}{N(1+\rho)} \sum_{a} \hat\eta_{_M} z_{1}^{(a)} \sigma^{(a)}+J\SOMMA{a}{}\xi^1\sigma^{(a)}\right)\right].
\end{align}
and $\hat{\eta}_{M}=(rM)^{-1}\sum_{A=1}^M\eta^{1,A}$.
Assuming (as usual within the replica method) that the limits $N\to\infty$ and $n\to 0$ 
can be interchanged, the integrals can be performed by steepest descent. This leads to
\begin{align}
    -\b \mathcal{F}^{(P)}(J)= \lim_{n\to 0} \dfrac{1}{n} A[\bm Q^\star, \bm P^\star, \bm Z^\star, \bm N ^\star] 
    \label{eq:trick_relation_sup}
\end{align}
where $\bm Q^\star, \bm P^\star, \bm Z^\star, \bm N ^\star$, are solution of the saddle point equations 
\begin{eqnarray}
&\dfrac{\partial A}{\partial p_{ab}}=0&\quad\hence\quad 
q_{ab}^*=\langle \sigma^a \sigma^b\rangle_{\rm eff}
\label{eq:selfQ_sup}
\\
&\dfrac{\partial A}{\partial z_{1}^{(a)}}=0&\quad\hence\quad 
(n_{1}^{(a)})^*=\dfrac{r}{\R}\SOMMA{A=1}{M}\mathbb{E}\langle \eta^{1,A}\sigma^a \rangle_{\rm eff}
\label{eq:selfm_sup}
\\
&\dfrac{\partial A}{\partial q_{ab}}=0&\quad\hence\quad 
p_{ab}^*=i \b\,^2 \dfrac{P}{4}\gamma N (q_{ab}^*)^{P-1}
\label{eq:selfP_sup}
\\
&\dfrac{\partial A}{\partial n_{1}^{(a)}}=0&\quad\hence\quad 
(z_{1}^{(a)})^*=i \b \dfrac{P}{2} (1+\rho)^{P/2} N [(n_{1}^{(a)})^*]^{P-1}
\label{eq:selfZ_sup}
\end{eqnarray}
which provides a set of self-consistency equations for $\bm Q$, $\bm N$, $\bm P$, and $\bm Z$. Using the last two equations to eliminate $\bm P$ and $\bm Z$ from the description, we finally obtain
\begin{align}
    -\b \mathcal F(J)&=\lim_{n\to 0} \dfrac{1}{n} \left\{\dfrac{1}{4}{\b}^2\gamma (P-1)\SOMMA{a,b}{n,n}[q_{ab}^*]^P + \dfrac{\b (1+\rho)^{P/2}}{2}(P-1) \sum_{a} [(n_{1}^{(a)})^*]^P \notag \right.
    \\
    & \left.-\mathbb{E} \log \left[ \sum_{\bm \sigma} \exp\left( \b\,^2 \dfrac{P}{4}\gamma\sum_{a,b}    [q_{ab}^*]^{P-1} \sigma^{(a)}\sigma^{(b)} + \b \dfrac{P}{2} (1+\rho)^{P/2-1}\sum_{a} \hat\eta_{_M}     [(n_{1}^{(a)})^*]^{P-1} \sigma^{(a)}\right.\right.\right. \notag \\ 
    &\left. \left. \left. +J\SOMMA{a}{}\xi^1\sigma^{(a)}\right)\right]\right\}\,.
    \label{eq:rt_free_sup}
\end{align}

To make progress, we need to find the form of $\bQ$ and $\bm N$ in the limit $n\to 0$. Again, we use the $1$RSB ansatz for the two-replica overlap provided in \eqref{eq:qabassump}, while we assume that $n_{1,a}$ is self-averaging, i.e. $n_{1}^{(a)}= \n$.
Proceeding analogously to the unsupervised case, after taking the limit $n\to 0$ we find
the same expression for the quenched free energy (see \eqref{eq:pressure_Guerra_sup}) and order parameters (see \ref{eq:High_store_self_n_q_sup}) previously obtained by using Guerra's interpolation technique. 
 
\subsection{Limiting cases: $M\to \infty $ and $\beta \to \infty$}
Now we state the next corollaries concerning the large datasets $M \to\infty$ limit and the ground state  $\beta \to \infty$ limit. We omit their proofs since these are trivial variations of those provided for the unsupervised case (cfr. Corollaries \ref{cor:large} and  \ref{cor:nulltemp_rho}).

\begin{corollary}
\label{cor:large_sup}
In the large dataset limit $M\to \infty$, the 1RSB self-consistency equations for the order parameters of the supervised DAM model can be expressed as
\begin{align}
    \n&= \dfrac{\m}{1+\rho} +\beta ' \dfrac{P}{2}\rho(1+\rho)^{P/2-2}\left[1+(\theta-1)\q_2-\theta \q_1\right]\n^{^{P-1}},\notag
    \\
    \q_1&= \mathbb{E}_{Y^{(1)}} \left[\dfrac{\mathbb{E}_{Y^{(2)}} \cosh^\theta g(\beta, \gamma, \bm Y) \tanh g(\beta, \gamma, \bm Y)}{\mathbb{E}_{Y^{(2)}} \cosh^\theta g(\beta, \gamma, \bm Y)} \right]^2, \notag
         \\
    \q_2 &= \mathbb{E}_{Y^{(1)}} \left[\dfrac{\mathbb{E}_{Y^{(2)}} \cosh^\theta g(\beta, \gamma, \bm Y) \tanh^2 g(\beta, \gamma, \bm Y)}{\mathbb{E}_{Y^{(2)}} \cosh^\theta g(\beta, \gamma, \bm Y)} \right],
    \label{eq:selfLarge_sup}
    \\
    \m&= \mathbb{E}_{_1}\left[\dfrac{\mathbb{E}_{Y^{(2)}} \cosh^\theta g(\beta, \gamma, \bm Y) \tanh g(\beta, \gamma, \bm Y)}{\mathbb{E}_{Y^{(2)}} \cosh^\theta g(\b, \gamma, \bm Y)}\right].\notag
\end{align} 
where  the expression of $g$ reads as
 \begin{align}
    g(\beta, \gamma, \bm Y)=&\beta '\dfrac{P}{2}\n^{^{P-1}}(1+\rho)^{P/2-1} + Y^{(2)} \b\sqrt{
             \gamma\dfrac{  P}{2} (\q_2^{^{P-1}} - \q_1 ^{^{P-1}})} \notag \\
             &+  Y^{(1)}\b\sqrt{\rho\left(\dfrac{P}{2}\n^{^{P-1}}(1+\rho)^{P/2-1}\right)^2+  \gamma
    \dfrac{P}{2}\,\q_1^{^{P-1}}\;}\;.
    \label{eq:g_of_unsuper_n_app_sup}
 \end{align}
 \normalsize

Moreover, if we use the truncated expression for $\n$, namely
\begin{equation}
    \n = \dfrac{\m}{1+\rho}
\end{equation}
we get the simplified expression of \eqref{eq:g_of_unsuper_n_app_sup} that is
 \begin{align}
    g(\beta, \gamma, \bm Y)=&\tilde\beta\dfrac{P}{2}\m^{^{P-1}} + Y^{(2)} \tilde\beta \sqrt{
             \gamma\dfrac{P}{2}(1+\rho)^P (\q_2^{^{P-1}} - \q_1 ^{^{P-1}})} \notag \\
             &+  Y^{(1)}\tilde\beta\sqrt{\rho\left(\dfrac{P}{2}\m^{^{P-1}}\right)^2+  \gamma(1+\rho)^P
    \dfrac{P}{2}\,\q_1^{^{P-1}}\;}\;,
    \label{eq:g_of_unsuper_n_app_sup_troncata}
 \end{align}
 where $\tilde\beta=\b/(1+\rho)^{P/2}$.
\end{corollary}


\begin{figure}[t]
    \centering
    \includegraphics[width = 15cm]{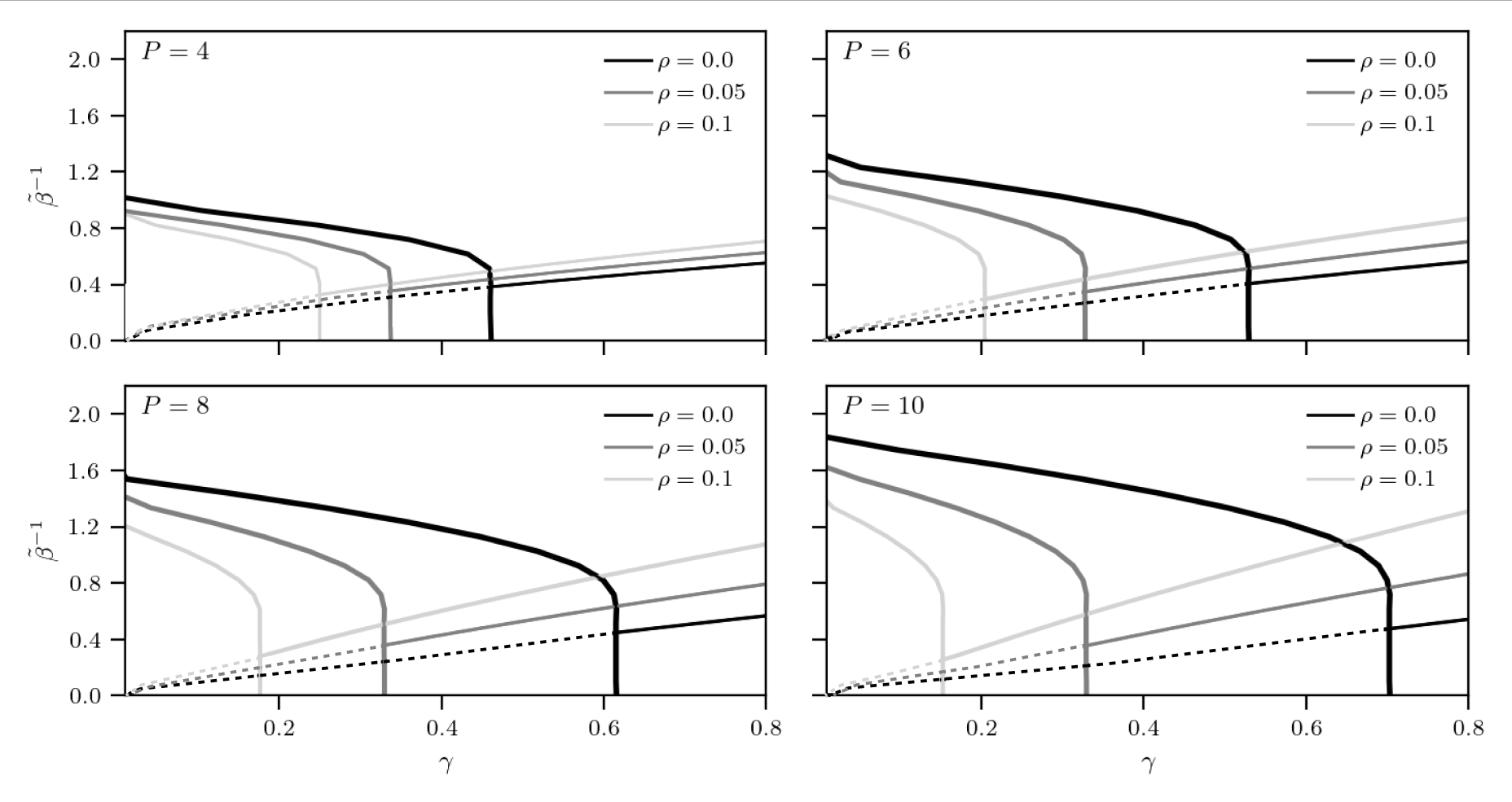}
    \caption{Phase Diagram of the supervised DAM model for different values of the dataset entropy $\rho$, ranging from $\rho=0.1$ (light grey stars), $0.05$ (grey triangles) up to $0$ (black dots). The interpretation of the different regions is the same as in the unsupervised model. As $\rho$ decrease (i.e. as more information is provided to the network), the instability region shrinks and the retrieval region expands, similarly to what was observed earlier upon increasing $M$.}
    \label{fig:sup}
\end{figure}

\begin{corollary}
The ground-state (i.e. $\beta \to \infty$) self-consistency equations for the order parameters of the theory in
the large dataset limit (i.e. $M\gg 1$) and under the $1$RSB assumption are
\begin{align}
    \bar{n} &= \dfrac{\m}{(1+\rho)}  \\
    \m &= 1- 2 \mathbb{E}_{Y^{(1)}} \left[ \dfrac{1}{1+\exp\left( 2D(A_1 + A_2 Y^{(1)})\right)\mathcal{Q}(A)}\right] \label{eq:m_beta}\\
    \Delta \q&=\q_2-\q_1 = 4 \mathbb{E}_{Y^{(1)}} \left\{ \dfrac{\exp\left( 2D(A_1 + A_2 Y^{(1)})\right)\mathcal{Q}(A)}{\left[1+\exp\left( 2D(A_1 + A_2 Y^{(1)})\right)\mathcal{Q}(A)\right]^2}\right\} \label{eq:q_beta_sup}
\end{align}
where $D=\tilde\beta \theta$ and
\begin{equation}
    \begin{array}{lll}
    &\mathcal{Q}(A)= \dfrac{1+\mathrm{erf}(K^{+})}{1+\mathrm{erf}(K^{-})}, \ \ \ &K^{\pm}= \dfrac{D A_3^2 \pm (A_1 + A_2 Y^{(1)})}{A_3 \sqrt{2}}, 
    \\\\
    &A_1 = \dfrac{P}{2}\m^{P-1}, \ \ \ & A_2 = \sqrt{\rho\left(\dfrac{P \m^{P-1}}{2}\right)^2+\gamma(1+\rho)^P\dfrac{P}{2}},
    \\\\
    &A_3 = \sqrt{\gamma(1+\rho)^P\dfrac{P(P-1)}{2}\Delta\q}.
    \end{array}
\end{equation}
\end{corollary}
By numerically solving the self-consistency equations \ref{eq:selfLarge_sup}, we obtain the phase diagram shown in Fig. \ref{fig:sup}, for different values of $P$ and different values of the dataset entropy $\rho$, as shown in the legend.

\begin{figure}[t]
    \centering
    \includegraphics[width=14cm]{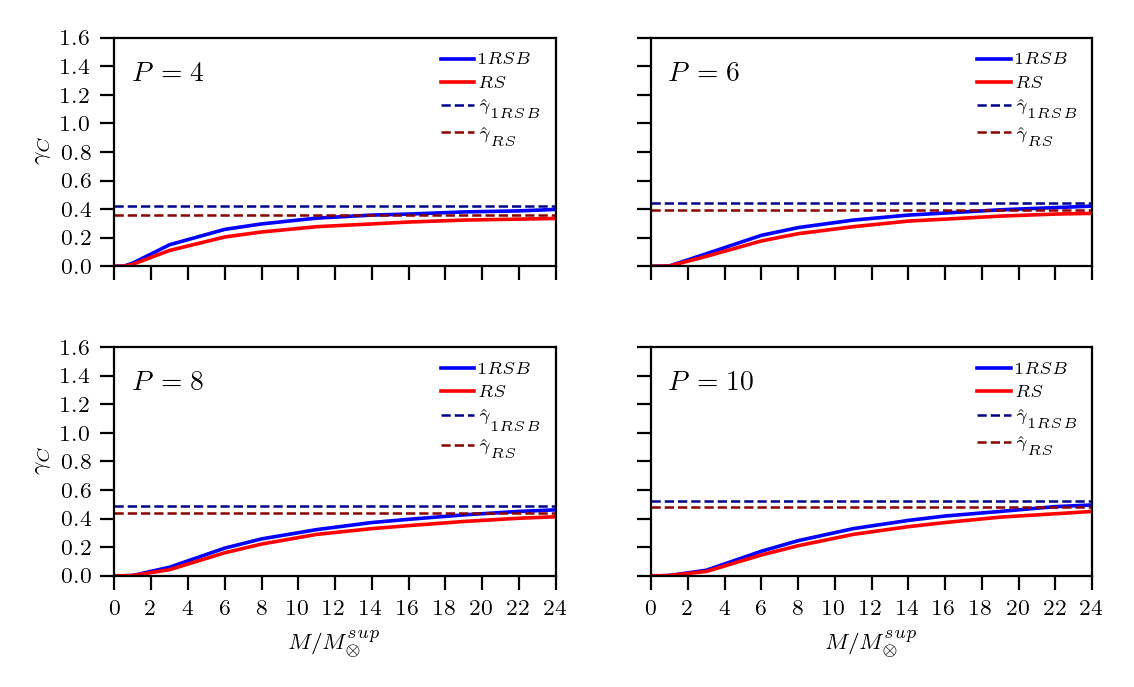}
    \caption{Critical storage capacity $\gamma_c$ in the ground state for the supervised DAM model for example noise $r=0.2$, as a function of the ratio $M/M_\otimes^{sup}$, both for the RS (red line) and $1$RSB (blue line) assumptions. Different panels show results for different values of $P=4,\ 6,\ 8,\ 10$.}
    \label{fig:null_T_superv}
\end{figure}


In \cite{super} it has been proven that the minimum number of examples $M_{\otimes}(r, \gamma)$ required by the supervised DAM model to correctly learn and retrieve an archetype, is given by 
\begin{align}
   M_{\otimes}(r, \gamma)= \left(1+2\gamma\right)\dfrac{1}{r^{\2}}
\end{align}
where $\gamma \neq 0$. 
In Fig. \ref{fig:null_T_superv} we plot the critical storage $\gamma_C$ in the ground state versus the ratio $M/M_\otimes$. As previously observed for the unsupervised setting, the phenomenon of replica symmetry breaking slightly increases the critical storage and it does not alter the minimum number of examples required for learning.


\section{Conclusions and outlooks}
\label{sec:conclusion}

This manuscript analyses the equilibrium behaviour of Dense Associative Memories (DAM) trained with or without the supervision of a teacher, within a 
$1$RSB assumption, thus extending previous analysis carried out at a RS level \cite{super,unsup}.
The unsupervised and supervised settings differ 
in the choice of the couplings (which involve $P$ nodes) and are given by
\begin{align}
    J^{unsup}_{i_1i_2...i_P} &\sim \frac{1}{ M N^{P-1}} \sum_{\mu=1}^K \sum_{a=1}^M \eta_{i_1}^{\mu,a} ...\:\eta_{i_P}^{\mu,a}, \\
    J^{sup}_{i_1i_2...i_P} &\sim \frac{1}{ M^P N^{P-1}} \sum_{\mu=1}^K \sum_{a_1, \hdots, a_P=1}^{M, \hdots, M} \eta_{i_1}^{\mu,a_1} ...\:\eta_{i_P}^{\mu,a_P}
\end{align}
respectively, 
where $\left\lbrace\boldsymbol \eta^{\mu}_a \right\rbrace_{a=1,...,M}^{\mu=1,...,K}$ are perturbed versions of the unknown archetypes $\left\lbrace \bm\xi^\mu\right\rbrace^{\mu=1, \hdots , K}$. 
The network does not experience the archetypes directly, instead it has to infer them from the supplied examples.
For both the settings, we obtained explicit expressions for the quenched free energy and derived full phase diagrams.
In doing so, we proved a full equivalence, at $1$RSB level, between two different approaches, namely 
Guerra's telescopic interpolation  \cite{guerra_broken} and Parisi's RSB theory \cite{MPV}. In addition, we derived (in Appendix \ref{Sec:labile}) the De Almeida-Thouless line, which marks the onset of the instability of the replica symmetric description, and below which the $1$RSB description should be preferred. 

The main differences brought about by the RSB description, with respect to the RS one, consist in the disappearance of the instability region within the retrieval zone of the phase diagram close to saturation (as standard in glassy statistical mechanics \cite{Amit}) and in a slight improvement of the value of the critical storage. 
Importantly, the threshold for learning, both in the supervised and unsupervised settings, is not influenced by replica symmetry breaking, i.e. the minimum number of examples required to infer the archetypes is the same in the RS and 1RSB description. 
Interestingly, the optimal value of the Parisi's parameter $\theta$, that controls the distribution of the overlaps in the $1$RSB scenario, is not influenced by the dataset entropy and is equal to that of the classical Hopfield model. 
From the mathematical viewpoint, possible future developments would be  relaxing the constraints of a self-averaging Mattis magnetization and inspecting how the learning and retrieval properties of these networks change with  different kind of noise: in this work we have focused on multiplicative noise, however the use of additive noise has lately gained large popularity in generative models for machine learning (see for instance \cite{MarcGiulio2023}). Furthermore, within the framework of multiplicative noise, an interesting outlook would be considering corruptions of  archetypes which also consist of blank (in addition to inverted) entries, as done recently in \cite{AdriFede2023}  for networks away from saturation. Their operation in the saturated regime and the effects of replica symmetry breaking have not yet been investigated: we plan to report soon on these topics. 

\section*{Acknowledgements}

All the authors acknowledge the stimulating research environment provided by the Alan Turing Institute’s Theory and Methods Challenge Fortnights event {\em Physics-informed Machine Learning}.
Albanese acknowledges Ermenegildo Zegna Founder's Scholarship, UMI (Unione Matematica Italiana), INdAM – GNFM Project (CUP E53C22001930001) and PRIN grant {\em Stochastic Methods for Complex Systems} n. 2017JFFHS for financial support and King's College London for kind hospitality. 
Alessandrelli acknowledges INdAM (Istituto Nazionale d'Alta Matematica) and Unisalento for support via PhD-AI.  
Barra's research is supported by MAECI via the BULBUL grant (Italy-Israel collaboration), Project n. F85F21006230001 {\em  Brain-inspired Ultra-fast and Ultra-sharp machines for assisted healthcare} and by MUR via the PRIN-2022 grant {\em Statistical Mechanics of Learning Machines: from algorithmic and information-theoretical limits to new biologically inspired paradigms}, Project n. 20229T9EAT that are gratefully acknowledged. 

\appendix

\section{Instability of the RS solution: AT lines}\label{Sec:labile}

In the main text we 
have analysed the equilibrium behaviour of supervised and unsupervised DAM models 
under the assumption of replica symmetry breaking. While such phenomenon is 
expected in these models, a formal proof that the replica symmetric theory becomes unstable in certain ranges of the control parameters has not been provided in the literature. 
In this section we provide such a proof and we derive the critical line of the RS instability 
in the phase diagram, for both the unsupervised and the supervised DAM models, separately. We will use the method recently introduced in \cite{albanese2023almeida}, which 
provides a simple alternative to the method 
originally introduced by de Almeida and Thouless 
in \cite{de1978stability}, as it does not require to compute 
the so-called replicon (i.e. the smallest eigenvalue of the spectrum) of the Hessian of the quadratic fluctuations of the free-energy around its RS value and it does not rely on the availability of an ``ansatz-free" expression for the free-energy.

\subsection{AT line for DAM in unsupervised setting}
\label{sec:AT_unsup}

Following the method introduced in \cite{albanese2023almeida}, we aim to 
determine the region in the phase diagram where the quenched free energy evaluated within the $1$RSB approximation, that from now on we  denote for convenience as $\mathcal F^{(P)}_{1RSB}(\n, \q_2,\q_1|\theta)$, is smaller than  the free energy evaluated within the RS assumption, $\mathcal F^{(P)}_{RS}(\nRS, \q)$, 
in the limit $\theta \to 1$, where the transition from RS to RSB is expected to occur
.

We start by recalling the expression for 
$\mathcal F^{(P)}_{1RSB}(\n, \q_2,\q_1|\theta)$, given in \eqref{eq:pressure_GuerraRS_finale}, with $\n$, $\q_1$ and $\q_2$ determined from the self-consistency equations \eqref{eq:High_store_self_n_q}, and by 
providing the expression for the quenched free energy within the RS assumption, as derived in \cite{unsup} 
\begin{align}
    -\b \mathcal F^{(P)}_{RS}(\nRS,\q ) =& - \dfrac{{\b}^2 (1+\rho)^{P/2}}{2}(P-1) \nRS^P + \dfrac{{\b}^2 \gamma (1+\sqrt{\rho_P})^2}{4(1+\rho)^P}(1-\q^P)\notag \\
    &- \dfrac{{\b}^2 \gamma (1+\sqrt{\rho_P})^2}{4(1+\rho)^P} \q^{P-1} (1-\q)+\ln 2 \notag \\
    & + \mathbb{E} \ln \cosh \left( \b (1+\rho)^{P/2-1} \dfrac{P}{2} \etaM \nRS^{P-1} + \b z \sqrt{\gamma \dfrac{(1+\sqrt{\rho_P})^2}{(1+\rho)^P} \dfrac{P}{2} \q^{P-1}}\right) 
\end{align}
where $\mathbb{E}=\mathbb{E}_{\bm\xi^1}\mathbb{E}_{(\bm\eta^1|\bm\xi^1)}\mathbb{E}_z$, $\etaM:=\frac{1}{rM}\SOMMA{a=1}{M}\eta^{1,a}$, and $\nRS$ and $\q$ fulfill the following self-consistency equations
\begin{equation}
    \begin{array}{lll}
         \nRS=\dfrac{1}{1+\rho}\mathbb{E}\left\{\tanh{\left[\beta '\dfrac{P}{2}\nRS^{P-1}(1+\rho)^{P/2-1} \etaM+z \beta '\sqrt{ \gamma
         \dfrac{P}{2}\dfrac{(1+\sqrt{\rho_P})^2}{(1+\rho)^P } \q^{^{P-1}}}\;\right]}\hat{\eta}_M\right\}\,
    \end{array}
\end{equation}
\begin{equation}
    \begin{array}{lll}
         \q=\mathbb{E}\left\{\tanh^{\2}{\left[\beta '\dfrac{P}{2}\nRS^{P-1}(1+\rho)^{P/2-1} \etaM+z \beta '\sqrt{ \gamma
         \dfrac{P}{2}\dfrac{(1+\sqrt{\rho_P})^2}{(1+\rho)^P } \q^{^{P-1}}}\;\right]}\right\}.
    \end{array}
    \label{eq:self_unsup}
\end{equation}
%
\new{We note that for $\theta=1$, $\bar{q}_1=\bar{q}$, $\n=\nRS$ and the 1RSB expression for the quenched free-
energy reduces to the RS one.
Now, we expand the $1$RSB expression for the quenched free energy $\mathcal F^{(P)}_{1RSB}(\n,\q_2,\q_1|\theta)$, as 
given in \eqref{eq:pressure_GuerraRS_finale}, around $\theta=1$, using  $\mathcal F^{(P)}_{1RSB}(\n,\q_2,\q_1|\theta)\vert_{\theta=1}=\mathcal F^{(P)}_{RS}(\nRS,\q)$} 
\begin{align}
\label{eq:expansion}
    \mathcal{F}_{1RSB}(\n,\q_2,\q_1|\theta)= \mathcal{F}_{RS}(\nRS,\q)+ (\theta -1)\partial_\theta \mathcal{F}_{1RSB} (\n,\q_2,\q_1|\theta)\vert_{\theta=1}.
\end{align}
Since $\q_1$ and $\q_2$ are determined through the self-consistency equations  
\eqref{eq:High_store_self_n_q}, they depend on $\theta$ as well, so we have to expand \eqref{eq:High_store_self_n_q}
 around $\theta=1$ too. Following Ref. \cite{albanese2023almeida} closely, we obtain
\begin{eqnarray}
\q_1&=& \q+(\theta-1) A( \nRS,\, \q,\qqq(\nRS,\, \q)) \label{eq:qq_t1}\\
        \q_2&=& \qqq(\nRS,\q)+(\theta-1) B( \nRS,\, \q,\qqq(\nRS,\, \q))
        \label{eq:q1-q_t1}
\end{eqnarray}
where $\q$ is the solution of \eqref{eq:self_unsup}, $\qqq(\nRS,\q)$ is the solution of the following self-consistency equation 
\begin{align}
    \q_2= \mathbb{E}_{Y^{(1)}} \left[ \dfrac{\mathbb{E}_{Y^{(2)}} \cosh {g(\nRS,\q,\qqq(\nRS,\, \q))} \tanh^2 {g(\nRS,\q,\qqq(\nRS,\, \q))}}{\mathbb{E}_{Y^{(2)}} \cosh {g(\nRS,\q,\qqq(\nRS,\, \q))}}\right] 
    \label{eq:self2_t1}
\end{align}
where
\begin{align}
    g(\nRS,\qqq(\nRS,\q), \q) =& \beta '\dfrac{P}{2}\nRS^{P-1}(1+\rho)^{P/2-1} \etaM+Y^{(1)} \beta '\sqrt{\gamma\dfrac{(1+\sqrt{\rho_P})^2}{(1+\rho)^{P}}  \dfrac{P}{2}\q^{^{P-1}}} \notag \\
    &+Y^{(2)} \beta '\sqrt{
    \gamma\dfrac{(1+\sqrt{\rho_P})^2}{(1+\rho)^{P}}\dfrac{P}{2} \left((\qqq(\nRS,\q))^{^{P-1}} -  \q ^{^{P-1}}\right)}.
    \label{eq:g_AT}
\end{align}
and the functions $A(\nRS,\qqq(\nRS,\q), \q)$ and $B(\nRS,\qqq(\nRS,\q), \q)$ are given by
\begin{align}
A(\nRS,\qqq(\nRS,\q), \q)=&2\mathbb{E}_{Y^{(1)}}\left\{\dfrac{\mathbb{E}_{Y^{(2)}} \ln \cosh g(\nRS,\qqq(\nRS,\q), \q) \sinh g(\nRS,\qqq(\nRS,\q), \q) \tanh g(\nRS,\qqq(\nRS,\q), \q)}{\mathbb{E}_{Y^{(2)}} \cosh g(\nRS,\qqq(\nRS,\q), \q) }\right\} \notag \\
&-2\mathbb{E}_{Y^{(1)}} \left\{\dfrac{\mathbb{E}_{Y^{(2)}} \ln \cosh g(\nRS,\qqq(\nRS,\q), \q) \cosh g(\nRS,\qqq(\nRS,\q), \q)}{\left(\mathbb{E}_{Y^{(2)}} \cosh g(\nRS,\qqq(\nRS,\q), \q) \right)^2} \right. \notag \\
&\left. \hspace{1.5cm} \cdot \dfrac{\mathbb{E}_{Y^{(2)}} \sinh g(\nRS,\qqq(\nRS,\q), \q) \tanh g(\nRS,\qqq(\nRS,\q), \q)}{\left(\mathbb{E}_{Y^{(2)}} \cosh g(\nRS,\qqq(\nRS,\q), \q) \right)^2 }\right\}\,,
    \label{eq:A-Dense_t1}
    \end{align}
    \begin{align}
B(\nRS,\qqq(\nRS,\q), \q)=& 2\mathbb{E}_{Y^{(1)}} \left\{ \dfrac{\mathbb{E}_{Y^{(2)}} \sinh g(\nRS,\qqq(\nRS,\q), \q) \tanh  g(\nRS,\qqq(\nRS,\q), \q)  }{\left( \mathbb{E}_{Y^{(2)}} \cosh  g(\nRS,\qqq(\nRS,\q), \q)\right)^2} \right. \notag \\
&\hspace{1.5cm}\left.\cdot \dfrac{\mathbb{E}_{Y^{(2)}} \cosh g(\nRS,\qqq(\nRS,\q), \q) \log \cosh  g(\nRS,\qqq(\nRS,\q), \q) }{\left( \mathbb{E}_{Y^{(2)}} \cosh  g(\nRS,\qqq(\nRS,\q), \q)\right)^2}\right\}\notag \\
&-2\mathbb{E}_{Y^{(1)}} \left\{ \dfrac{\left(\mathbb{E}_{Y^{(2)}} \sinh g(\nRS,\qqq(\nRS,\q), \q)\tanh  g(\nRS,\qqq(\nRS,\q), \q) \right)^2 }{\left( \mathbb{E}_{Y^{(2)}} \cosh  g(\nRS,\qqq(\nRS,\q), \q) \right)^3}\right. \notag \\ 
&\hspace{1.5cm}\left. \cdot \dfrac{\mathbb{E}_{Y^{(2)}} \sinh g(\nRS,\qqq(\nRS,\q), \q) \log \cosh  g(\nRS,\qqq(\nRS,\q), \q) }{\left( \mathbb{E}_{Y^{(2)}} \cosh  g(\nRS,\qqq(\nRS,\q), \q) \right)^3}\right\},
    \label{eq:B-Dense_t1}
    \end{align}
respectively. Similarly, we can expand $\n$ as
\begin{equation}
    \n = \nRS + (\theta-1) C( \nRS,\, \q,\qqq(\nRS,\, \q))
    \label{eq:m_t1}
\end{equation}
where
\begin{align}
\begin{array}{lll}
     &C(\nRS, \qqq(\nRS,\q),\q)=\mathbb{E}_{Y^{(1)}} \left[ \dfrac{\mathbb{E}_{Y^{(2)}} \sinh g(\nRS,\qqq(\nRS,\q), \q)\log \cosh g(\nRS,\qqq(\nRS,\q), \q)}{\mathbb{E}_{Y^{(2)}} \cosh g(\nRS,\qqq(\nRS,\q), \q) }\right]  
     \\\\
        & -\mathbb{E}_{Y^{(1)}} \left[ \dfrac{\mathbb{E}_{Y^{(2)}} \sinh g(\nRS,\qqq(\nRS,\q), \q)\mathbb{E}_{Y^{(2)}} \cosh g(\nRS,\qqq(\nRS,\q), \q) \log \cosh g(\nRS,\qqq(\nRS,\q), \q)}{\left(\mathbb{E}_{Y^{(2)}} \cosh g(\nRS,\qqq(\nRS,\q), \q)\right)^2 }\right]
\end{array}
\label{eq:C-Dense_t1}
\end{align}
In the limit $\theta \to 1$, Eq. \eqref{eq:expansion} implies that $\mathcal F^{(P)}_{1RSB}(\nRS,\q_2(\q), \q|\theta)< \mathcal F^{(P)}_{RS}(\nRS,\q)$ when we have that $\partial_\theta \mathcal{F}_{1RSB} (\nRS,\q_2(\q), \q|\theta)\vert_{\theta=1}>0$.
Next, we evaluate 
\begin{align}
\label{eq:KDense_t1}
    K(\nRS, \qqq(\nRS,\q), \q)&\coloneqq \partial_\theta (-\b \mathcal F^{(P)}_{\rm 1RSB}(\nRS,\q_2(\q), \q|\theta) )\vert_{\theta=1} 
    \notag\\
    =&-\dfrac{{\b}^2 \gamma (1+\sqrt{\rho_P})^2}{4(1+\rho)^P} (P-1)[(\tilde{q}_2(\q))^P- \q^P] -\dfrac{{\b}^2 \gamma (1+\sqrt{\rho_P})^2}{4(1+\rho)^P} P[(\tilde{q}_2(\q))^{P-1} - \q^{P-1}] \notag \\
    &- \mathbb{E} \ln \cosh \left( \beta '\dfrac{P}{2}\nRS^{P-1}(1+\rho)^{P/2-1} \etaM+Y \beta '\sqrt{ \gamma
         \dfrac{P}{2}\dfrac{(1+\sqrt{\rho_P})^2}{(1+\rho)^P } \q^{^{P-1}}}\right) \notag \\
    &+ \mathbb{E}_{Y^{(1)}} \left[ \dfrac{\mathbb{E}_{Y^{(2)}} \cosh g(\nRS,\qqq(\nRS,\q), \q) \log \cosh g(\nRS,\qqq(\nRS,\q), \q) }{\mathbb{E}_{Y^{(2)}} \cosh g(\nRS,\qqq(\nRS,\q), \q) }\right].
\end{align}
In order to determine the sign of the expression above, it is useful to note that $K(\nRS,\q, \q)=0$, as the last term of \eqref{eq:g_AT} vanishes so, the last two addends in \eqref{eq:KDense_t1} elide each other. This is as expected, as $\q$ is an extremum of the RS free-energy, which is retrieved for $\theta=1$. Next, we study $K(\nRS,x, \q)$ as a function of $x \in [0, \q]$ and locate its extrema. These are found from 
\begin{align}
    \partial_{x} K(\nRS, x, \q)=& -\dfrac{{\b}^2 \gamma (1+\sqrt{\rho_P})^2 (P-1) P x^{P-2}}{4(1+\rho)^P}\notag \\
    &\hspace{0.5cm}\cdot\left[x- \mathbb{E}_{Y^{(1)}} \left[ \dfrac{\mathbb{E}_{Y^{(2)}} \sinh g(\nRS,\q, x) \tanh g(\nRS,\q, x) }{\mathbb{E}_{Y^{(2)}} \cosh g(\nRS,\q, x) }\right]\right]=0
    \end{align}
as 
\begin{align} x = \mathbb{E}_{Y^{(1)}} \left[ \dfrac{\mathbb{E}_{Y^{(2)}} \sinh g(\nRS,\q, x) \tanh g(\nRS,\q, x) }{\mathbb{E}_{Y^{(2)}} \cosh g(\nRS,\q, x) }\right] \equiv \qqq(\nRS,\q).
\end{align}
where the last equality follows from algebraic manipulations of trigonometric functions.

\begin{figure}
     \centering
     \includegraphics[width=15cm]{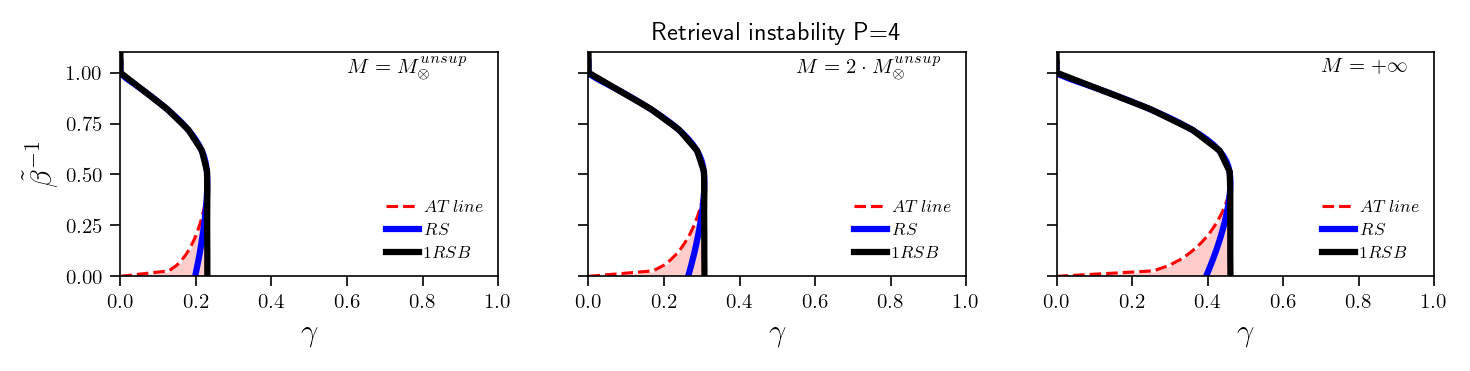}
     \\
     \includegraphics[width=15.5cm]{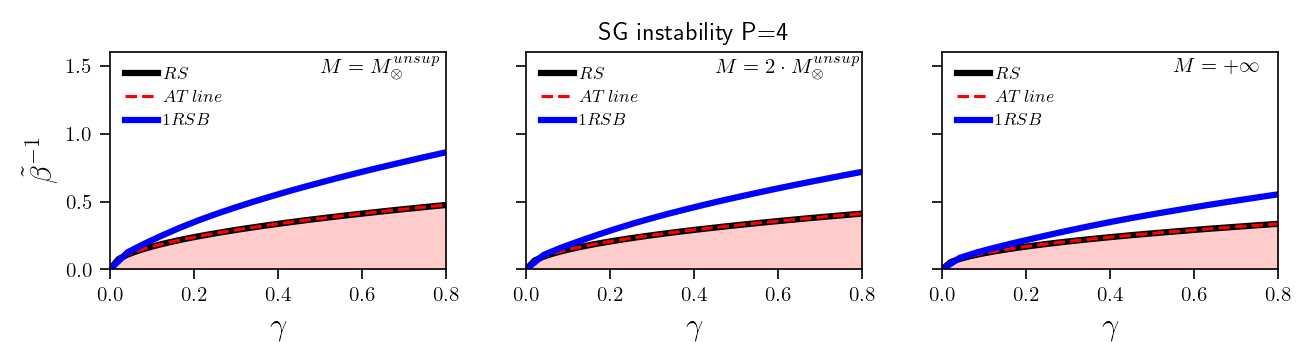}
     \caption{The AT line (marking the onset of  instability of the RS theory) is shown as a dashed line in the parameter space of  storage load $\gamma$ and scaled noise $\tilde \beta=\beta'/(1+\rho)^{P/2}$  for the {unsupervised} DAM model, with $P=4$.
     For comparison, we also show the critical lines delimiting the retrieval region (top plots) and the spin-glass region (bottom plots), within the RS and the RSB theory,  respectively, as obtained in the phase diagram in Fig. \ref{fig:unsup}. 
     Different columns correspond to different numbers of examples, $M$, presented to the network, namely $M=M_{\otimes}^{unsup}$ (left), $M=2M_{\otimes}^{unsup}$ (middle) and $M=+\infty$ (right). {We note that the AT line coincides with the line delimiting the SG region withing the RS theory.} 
     }
     \label{fig:AT line}
 \end{figure}

Given that $K(\nRS,x,\q)$ vanishes for $x=\q$, if the extremum $x=\qqq(\nRS,\q)$ is global in the domain considered, we must have that $K(\nRS,\tilde q_2(\nRS,\q), \q)>0$ if $x=\tilde q_2(\nRS,\q)$ is a maximum and $K(\nRS,\tilde q_2(\nRS,\q),\bar q)<0$ if $x=\tilde q_2(\nRS,\q)$ is a minimum. Evaluating 
\begin{align}
    \partial^2_{x} K(\nRS,x, \q)\vert_{x=\qqq(\nRS,\q)}=& -\dfrac{{\b}^2 \gamma (1+\sqrt{\rho_P})^2 (P-1) P}{4 (1+\rho)^P}  (\tilde{q}_2(\nRS,\q))^{P-2} \notag \\
    &\hspace{-1cm}\cdot\left[ 1-\dfrac{{\b}^2 \gamma (1+\sqrt{\rho_P})^2}{2(1+\rho)^P}(P-1) P (\tilde{q}_2(\nRS,\q))^{P-2} \mathbb{E}_{Y^{(1)}} \left[ \dfrac{\mathbb{E}_{Y^{(2)}} \textnormal{sech}^3 g(\nRS,\qqq(\nRS,\q), \q)  }{\mathbb{E}_{Y^{(2)}} \cosh g(\nRS,\qqq(\nRS,\q), \q)}\right]\right]
    \label{eq:K2-Dense}
\end{align}
we have that if the expression in the square brackets is positive, namely if the parameter $\gamma {\b}^2$ satisfies
\begin{align}
    \dfrac{{\b}^2 \gamma (1+\sqrt{\rho_P})^2}{2(1+\rho)^P}(P-1) P (\tilde{q}_2(\nRS,\q))^{P-2} \mathbb{E}_{Y^{(1)}} \left[ \dfrac{\mathbb{E}_{Y^{(2)}} \textnormal{sech}^3 g(\nRS,\qqq(\nRS,\q), \q)  }{\mathbb{E}_{Y^{(2)}} \cosh g(\nRS,\qqq(\nRS,\q), \q)}\right] < 1,
    \label{eq:AT1}
\end{align}
$K(\nRS,\qqq(\nRS,\q), \q)<0$ and 
 $\mathcal F^{(P)}_{1RSB}(\nRS,\qqq(\nRS,\q), \q|\theta)-\mathcal F^{(P)}_{RS}(\nRS,\q)=-(\theta-1)K(\nRS,\qqq(\nRS,\q), \q)/\b<0$, hence the RS theory 
 is unstable. 
We plot the line \eqref{eq:AT1} in Figure \ref{fig:AT line}, for different values of $M$, in the parameter space $(\gamma, \tilde\beta)$, where $\tilde\beta = \beta'/(1+\rho)^{P/2}$, together with the critical lines delimiting the retrieval region (top plots) and the spin-glass region (bottom plots), within the RS and the RSB theory, respectively. 
We note that the above recovers the expression for the RS instability line found in  DAM models \cite{albanese2023almeida} in the limits $r \to 0$ or $M \to \infty$, where the values of $\rho_P$ and $\rho$ vanish. 

\subsection{AT line for DAM in supervised setting}\label{sec:ATlineUnsup}
In this section we derive the RS instability line for the supervised DAM model.
We start by providing the expression for the quenched free-energy in RS assumption  as derived in \cite{Gardner} for the standard Dense Hopfield Model.
\begin{align}
    -\b \mathcal F^{(P)}_{RS}(\nRS,\q) =& \ln 2 - \dfrac{{\b}^2}{2}(P-1) (1+\rho)^{P/2} \nRS^P + \dfrac{{\b}^2 \gamma}{4}(1-\q^P)- \dfrac{{\b}^2 \gamma P }{4} \q^{P-1} (1-\q)\notag \\
    &+ \mathbb{E}_\xi \mathbb{E}_{(\eta \vert \xi)} \mathbb{E}_z \ln \cosh \left( \b \dfrac{P}{2} (1+\rho)^{P/2-1} \hat{\eta}_M \n^{P-1} + \b z \sqrt{\gamma \dfrac{P}{2} \q^{P-1}}\right) 
\end{align}
with $\b:=2\beta/P!$ and where  $\q$ and $\n$ satisfy the self-consistency equations  
\begin{align}
    &\nRS= \dfrac{1}{1+\rho}\mathbb{E}_{\bm\xi^1}\mathbb{E}_{(\bm\eta^1|\bm\xi^1)} \mathbb{E}_z \left\{\tanh \left[ \b \dfrac{P}{2} (1+\rho)^{P/2-1} \hat{\eta}_M \nRS^{P-1} + \b z \sqrt{\gamma \dfrac{P}{2} \q^{P-1}}\right]\hat{\eta}_M\right\} , \notag \\
    &\q= \mathbb{E}_{\bm\xi^1}\mathbb{E}_{(\bm\eta^1|\bm\xi^1)} \mathbb{E}_z \left\{\tanh^2 \left[ \b \dfrac{P}{2} (1+\rho)^{P/2-1} \hat{\eta}_M \nRS^{P-1} + \b z \sqrt{\gamma \dfrac{P}{2} \q^{P-1}}\right]\right\} .
    \label{eq:self_RS_Dense}
\end{align}
We also recall that the quenched free-energy within the 1RSB approximation, $\mathcal F^{(P)}_{1RSB}(\n,\q_1, \q_2 \vert \theta)$, is 
as given in \eqref{eq:pressure_Guerra_sup}, 
%
where the order parameters $\q_2$, $\q_1$ and $\n$ satisfy the set of self-consistency equations provided in  \eqref{eq:High_store_self_n_q_sup}. 
We note that for $\theta=1$, $\q_1=\q$, $\n=\nRS$ and the 1RSB expression for the quenched free-energy reduces to the RS one.
%
Now, we expand, to the leading order in $\theta-1$, the 1RSB quenched free-energy around its RS expression, as shown in \eqref{eq:expansion}.
Since the self-consistency equations also depend on $\theta$, we need to expand them too. We can write $\q_1$ as in \eqref{eq:qq_t1}, with $A(\nRS,\q_2(\nRS,\q),\q)$ given in \eqref{eq:A-Dense_t1}, and $\q_2$ as given in \eqref{eq:q1-q_t1}, 
where $\qqq(\nRS,\q)$ is the solution of \eqref{eq:self2_t1}
and  $B(\nRS,\q_2(\nRS,\q),\q)$ is as given in \eqref{eq:B-Dense_t1} (we recall that now the expression of $g(\nRS,\q_2(\nRS,\q),\q )$ is as given in \eqref{eq:g_sup}).
With these expressions in hand, we can now compute the derivative of $\mathcal F^{(P)}_{\rm 1RSB}$ w.r.t. $\theta$ when $\theta=1$, as needed in \eqref{eq:expansion}
\begin{align}
\label{eq:KDense_t1_sup}
    K(\nRS,\qqq(\nRS,\q), \q):=&\partial_\theta (-\b \mathcal F^{(P)}_{\rm 1RSB}(\nRS,\q_2(\nRS,\q), \q|\theta) )\vert_{\theta=1} 
    \notag\\
    &=-\dfrac{{\b}^2 \gamma }{4} (P-1)[(\tilde{q}_2(\nRS,\q))^P- \q^P] -\dfrac{{\b}^2 \gamma }{4} P[(\tilde{q}_2(\nRS,\q))^{P-1} - \q^{P-1}] \notag \\
    &- \mathbb{E}_\xi \mathbb{E}_{(\eta \vert \xi)} \mathbb{E}_Y \ln \cosh \left( \beta '\dfrac{P}{2}\nRS^{P-1}(1+\rho)^{P/2-1} \etaM+Y \beta '\sqrt{ \gamma
         \dfrac{P}{2}\q^{^{P-1}}}\right) \notag \\
    &+ \mathbb{E}_\xi \mathbb{E}_{(\eta \vert \xi)} \mathbb{E}_{Y^{(1)}} \left[ \dfrac{\mathbb{E}_{Y^{(2)}} \cosh g(\nRS,\qqq(\nRS,\q), \q) \log \cosh g(\nRS,\qqq(\nRS,\q), \q) }{\mathbb{E}_{Y^{(2)}} \cosh g(\nRS,\qqq(\nRS,\q), \q) }\right].
\end{align}
Again, we have that $K(\nRS,\q, \q)=0$ (as $\q$ is the extremum of the RS free-energy). Next, we inspect the sign of $K(\nRS,\qqq(\nRS,\q), \q)$. To this purpose, we study $K(\nRS,x,\q)$ for $x \in [0, \q]$ and locate its extrema, which are found from 
\begin{align}
    \partial_{x} K(\nRS, x, \q)=& -\dfrac{{\b}^2 \gamma (P-1) P x^{P-2}}{4(1+\rho)^P}\left[x- \mathbb{E}_{Y^{(1)}} \left[ \dfrac{\mathbb{E}_{Y^{(2)}} \sinh g(\nRS,\q, x) \tanh g(\nRS,\q, x) }{\mathbb{E}_{Y^{(2)}} \cosh g(\nRS,\q, x) }\right]\right]=0
    \end{align}
    as 
\begin{align} x = \mathbb{E}_{Y^{(1)}} \left[ \dfrac{\mathbb{E}_{Y^{(2)}} \cosh {g(\nRS,x, \q)} \tanh^2 {g(\nRS,x, \q)}}{\mathbb{E}_{Y^{(2)}} \cosh {g(\nRS,x, \q)}}\right] \equiv \qqq(\nRS,\q)
\end{align}
where the last equality follows from \eqref{eq:self2_t1}. 
Under the assumption that 
the extremum $x=\qqq(\nRS,\q)$ is global in the domain considered, we have that 
$K(\nRS,\qqq(\nRS,\q), \q)>0$ if $x=\tilde q_2(\nRS,\q)$ is a maximum and $K(\nRS,\qqq(\nRS,\q), \q)<0$
if it is a minimum. In particular, if 
\begin{align}
    \partial^2_{x} K(\nRS,x,\q)\vert_{x=\qqq(\nRS,\q)}=& -\dfrac{{\b}^2 \gamma  (P-1) P (\qqq(\nRS,\q))^{P-2}}{4}\notag \\
    &\cdot\left[ 1-\dfrac{{\b}^2 \gamma }{2}(P-1) P (\tilde{q}_2(\nRS,\q))^{P-2} \mathbb{E}_{Y^{(1)}} \left[ \dfrac{\mathbb{E}_{Y^{(2)}} \textnormal{sech}^3 g(\nRS,\qqq(\nRS,\q), \q)  }{\mathbb{E}_{Y^{(2)}} \cosh g(\nRS,\qqq(\nRS,\q), \q)}\right]\right]
    \label{eq:K2-Dense1}
\end{align}
is positive, $K(\nRS,\qqq(\nRS,\q), \q)<0$ and $\mathcal F^{(P)}_{1RSB}(\nRS,\q_2(\nRS,\q),\q|\theta)< \mathcal F^{(P)}_{RS}(\nRS,\q)$. This happens when the expression in the curly brackets of the equation above is negative, i.e. when the parameter $\gamma {\b}^2$ satisfies the inequality
\begin{equation}\label{eq:result_PHOP}
\dfrac{{\b}^2 \gamma }{2}(P-1) P (\tilde{q}_2(\nRS,\q))^{P-2} \mathbb{E}_{Y^{(1)}} \left[ \dfrac{\mathbb{E}_{Y^{(2)}} \textnormal{sech}^3 g(\nRS,\qqq(\nRS,\q), \q)  }{\mathbb{E}_{Y^{(2)}} \cosh g(\nRS,\qqq(\nRS,\q), \q)}\right] >1.
\end{equation}
\new{
We note that \eqref{eq:result_PHOP} is functionally identical to the expression found in \cite{albanese2023almeida} for the RS instability line of the standard DAM model, the only difference being encoded in the term $g(\nRS,\qqq(\nRS,\q), \q)$, which is here defined differently, as it reflects the supervised protocol.
}
  {In particular, in the limit of $r\to 0$ or $M\to \infty$, where $\rho$ and $\rho_P$ vanish, \eqref{eq:result_PHOP} retrieves the 
RS instability line of standard DAM models, as obtained in \cite{albanese2023almeida}.}

\section{Proofs}
\subsection{Proof of Lemma \ref{lemma:der}}
\label{app:lemma}
Defining the shorthand $\mathcal{Z}_2(J,t)=\mathcal{Z}_2^{(P)}(\bm{\eta}^1,\bm\lambda, \bm Y;J,t)$, 
we start by computing the derivative of Eq. \eqref{hop_GuerraAction} w.r.t. $t$ 
\begin{equation}
\begin{array}{lll}
     -\b\partial_t \mathcal{F}^{(P)}_{N}(J,t) &=& \dfrac{1}{N} \mathbb{E}_0 \mathbb{E}_{\bm Y^{(1)}}\partial_t\left[  \ln \left [\mathbb{E}_{\bm Y^{(2)}}  \mathcal Z_2^\theta(J, t)  \right ]^{1/\theta} \right]=\dfrac{1}{N} \mathbb{E}_0 \mathbb{E}_{\bm Y^{(1)}}\left[ \dfrac{\mathbb{E}_{\bm Y^{(2)}}\partial_t \mathcal Z_2(J, t)^{\theta-1}\partial_t \mathcal Z_2(J, t)}{\mathbb{E}_{\bm Y^{(2)}}  \mathcal Z_2(J, t)^\theta} \right]
    \\\\
    &=&\dfrac{1}{N} \mathbb{E}_0 \mathbb{E}_{\bm Y^{(1)}}\mathbb{E}_{\bm Y^{(2)}}\left[\mathcal{W}_{2,t} \dfrac{\partial_t \mathcal Z_2(J, t)}{\mathcal Z_2(J, t)} \right]
\end{array}
\label{eq:def_t_deriv_RSB}
\end{equation}
where $\mathbb{E}_0=\mathbb{E}_{\bm\xi^1}\mathbb{E}_{(\bm\eta^1|\bm\xi^1)}\mathbb{E}_{\bm\lambda}$ and $\mathcal{W}_{2,t}$ is defined in \eqref{eq:W2}. 
Recalling the definition of $\mathcal Z_2(J, t)$ in \eqref{eq:B2} we have:
\begin{equation}
\begin{array}{lll}
     \partial_t \mathcal Z_2(J, t)&=& \SOMMA{\bm\sigma}{}\mathcal B_2(\boldsymbol{\sigma} ;J,  t ) \Bigg[\left(\dfrac{\beta 'N}{2M}\left(\dfrac{\R}{r^2} \right)^{P/2}\SOMMA{a=1}{M}n_{1,a}^{^P}(\bm\sigma)\right) - \psi N \SOMMA{a=1}{M} n_{1,a}(\bm\sigma) 
     \\\\
    & &\hspace*{-0.8cm}+\dfrac{1}{2}\dfrac{\beta 'P!(1+\sqrt{\rho_P})\sqrt{K}}{2\sqrt{t} (1+\rho)^{P/2} \,N^{P-1}}\SOMMA{{i_1 <\cdots < i_P}}{N,\cdots,N}\lambda_{i_1,\hdots, i_P}\sigma_{_{i_1}}\cdots\sigma_{_{i_{_{P}}}}  -\dfrac{1}{2\sqrt{1-t}}\SOMMA{b=1}{2}\left(A_b\SOMMA{i=1}{N}Y_i^{(b)}\sigma_i\right) \Bigg].
\end{array}
\end{equation}
where, in order to lighten the notation we have set $\mathcal{B}^{(P)}_2(\bm\sigma|\bm\eta^1,\bm\lambda,\bm Y;J,t)=\mathcal{B}_2(\bm\sigma;J,t)$.
Inserting this expression in \eqref{eq:def_t_deriv_RSB} and using the definition 
\eqref{omegaNKM} for the average of a single replica of the system over the generalised 
Boltzmann factor $\mathcal{B}_2(\bm\sigma;J,t)$,
we obtain 
\begin{align}
    &-\b\partial_t \mathcal{F}^{(P)}_{N}(J,t)=\mathbb E_0  \mathbb{E}_{\bm{Y}^{(1)}}  \mathbb{E}_{\bm{Y}^{(2)}} \Big\{\dfrac{\beta '}{2M}\left(\dfrac{\R}{r^2} \right)^{P/2}\mathcal{W}_{2,t}  \omega_t \left(\SOMMA{a=1}{M}n_{1,a}^{^P}(\bm\sigma)\right) - \psi \mathcal{W}_{2,t}  \omega_t\left(\SOMMA{a=1}{M} n_{1,a}(\bm\sigma) \right)\Big\} \notag 
    \\
    &+\dfrac{1}{2}\dfrac{\b P!(1+\sqrt{\rho_P}) \sqrt{K}}{2 \sqrt{t} (1+\rho)^{P/2} N^{P}}\SOMMA{{i_1 <\cdots < i_P}}{N,\cdots,N} \mathbb E_0  \mathbb{E}_{\bm{Y}^{(1)}}  \mathbb{E}_{\bm{Y}^{(2)}} \left\{\mathcal{W}_{2,t}\dfrac{1}{\mathcal Z_2(J,t)} \sum_{\bm \sigma}  \Big(\lambda_{i_1,\hdots,i_P}\sigma_{_{i_1}}\cdots\sigma_{_{i_{_{P}}}} \Big)\mathcal B_2(\boldsymbol{\sigma} ;J,  t ) \right\} \notag 
    \\
    &-\dfrac{1}{2N\sqrt{1-t}}\SOMMA{b=1}{2}  \mathbb E_0  \mathbb{E}_{\bm{Y}^{(1)}}  \mathbb{E}_{\bm{Y}^{(2)}} \left\{\mathcal{W}_{2,t}\dfrac{1}{\mathcal Z_2(J,t)} \sum_{\bm \sigma} \left(A_b\SOMMA{i=1}{N}Y_i^{(b)}\sigma_i\right)\mathcal B_2(\boldsymbol{\sigma} ;J,  t )\right\}.
    \end{align}
Next, denoting the combined average over 
the quenched disorder and the Boltzmann distribution as 
\begin{equation}
\label{eq:combined}
\langle \cdot \rangle=\mathbb{E}_0\mathbb{E}_{\bm{Y}^{(1)}}\mathbb{E}_{\bm{Y}^{(2)}}\left[\mathcal{W}_{2,t}\omega_t(\cdot)\right]
\end{equation}
and applying the Stein's lemma \eqref{eqn:gaussianrelation2Pspin} to the standard Gaussian variables $Y_i^{(b)}$ and $\lambda^{\mu}_{i_1,\hdots,i_P}$, we get 
\begin{align}
    -\b\partial_t \mathcal{F}^{(P)}_{N} =& \dfrac{\beta '}{2M}\left(\dfrac{\R}{r^2} \right)^{P/2}\SOMMA{a=1}{M} \langle n_{1,a}^P(\bm\sigma) \rangle_t - \psi \SOMMA{a=1}{M} \langle n_{1,a}(\bm\sigma) \rangle_t  \notag  
    \\
    &+{\dfrac{{\b}^2 K (1+\sqrt{\rho_P})^2}{ 8 (1+\rho)^{P} \,N^{2P-1}}}\SOMMA{i_{_1},\cdots, i_{_{P}}}{N,\cdots,N} \mathbb E_0  \mathbb{E}_{\bm{Y}^{(1)}}  \mathbb{E}_{\bm{Y}^{(2)}} \left(\mathcal{W}_{2,t} \omega_t(1) + (\theta-1) \mathcal{W}_{2,t} \omega_t^2( \sigma_{_{i_1}}\cdots\sigma_{_{i_{_{P}}}})\right) \notag 
    \\
    &-{\dfrac{{\b}^2P! K(1+\sqrt{\rho_P})^2}{ 8 (1+\rho)^{P} \,N^{2P-1}}}\SOMMA{i_{_1}<\cdots< i_{_{P}}}{N,\cdots,N} \mathbb E_0  \mathbb{E}_{\bm{Y}^{(1)}}  \theta \left[\mathbb{E}_{\bm{Y}^{(2)}} \left(\mathcal{W}_{2,t} \omega_t( \sigma_{_{i_1}}\cdots\sigma_{_{i_{_{P}}}})\right)\right]^2 \notag 
    \\
    &-\dfrac{(A_1)^2}{2}\SOMMA{i=1}{N} \mathbb E_0  \mathbb{E}_{\bm{Y}^{(1)}}  \mathbb{E}_{\bm{Y}^{(2)}} \left(1 + (\theta-1) \mathcal{W}_{2,t} \omega_t^2(\si)\right)-\dfrac{(A_1)^2}{2}\theta\SOMMA{i=1}{N} \mathbb E_0  \mathbb{E}_{\bm{Y}^{(1)}} \left[ \mathcal{W}_{2,t} \omega_t(\si)\right]^2 \notag \\
    &-\dfrac{(A_2)^2}{2}\SOMMA{i=1}{N} \mathbb E_0  \mathbb{E}_{\bm{Y}^{(1)}}  \mathbb{E}_{\bm{Y}^{(2)}} \left(1 + (\theta-1) \mathcal{W}_{2,t} \omega_t^2(\si)\right) .
\end{align}
Finally, we use the definitions \eqref{eq:media1}
and \eqref{eq:media2} and we arrive at the compact expression: 
\begin{align}
    -\b\partial_t \mathcal{F}^{(P)}_{N} =&  \dfrac{\beta '}{2M}\left(\dfrac{\R}{r^2} \right)^{P/2}\SOMMA{a=1}{M} \langle n_{1,a}^P(\bm\sigma) \rangle_t - \psi \SOMMA{a=1}{M} \langle n_{1,a}(\bm\sigma) \rangle_t  \notag  \\
    &+{\dfrac{{\b}^2(1+\sqrt{\rho_P})^2}{ 4 (1+\rho)^{P} }} \dfrac{P!K}{2N^{P-1}}\left[1+ (\theta-1) \l q_{12}^{P}(\bm\sigma^{(1)},\bm\sigma^{(2)}) \r_{t,2} - \theta \l q_{12}^{P}(\bm\sigma^{(1)},\bm\sigma^{(2)}) \r_{t,1}\right]\notag
    \\
    &-\dfrac{A_1^2}{2} \left[ 1+ (\theta-1) \l q_{12}(\bm\sigma^{(1)},\bm\sigma^{(2)}) \r_{t,2} - \theta \l q_{12}(\bm\sigma^{(1)},\bm\sigma^{(2)}) \r_{t,1} \right] \notag \\
    &-\dfrac{A_2^2}{2} \left[ 1+(\theta-1) \l q_{12}(\bm\sigma^{(1)},\bm\sigma^{(2)}) \r_{t,2} \right].
    \label{eq:compact}
\end{align}
Next, we take the thermodynamic limit $N\to \infty$, using assumption \ref{eq:seconda}.
By manipulating the expression for the moments $\langle q_{12}^P(\bm\sigma^{(1)},\bm\sigma^{(2)})\rangle_{t,a}$, with $a=1,2$, using Newton's binomial theorem 
\begin{equation}
 \begin{array}{lll}
     \langle q_{12}(\bm\sigma^{(1)},\bm\sigma^{(2)})^P \rangle_{t,a}  &=&
     \langle [(q_{12}(\bm\sigma^{(1)},\bm\sigma^{(2)})-\bar{q}_a+\bar{q}_a]^P \rangle_{t,a}=\SOMMA{k=0}{P} \begin{pmatrix}P\\k\end{pmatrix} \langle (q_{12}(\bm\sigma^{(1)},\bm\sigma^{(2)})-\bar{q}_a)^k \rangle_{t,a} \bar{q}_a^{P-k}
     \nonumber\\
      & &\hspace*{-1.5cm}=
     P\,\bar{q}_a^{P-1}\langle q_{12}(\bm\sigma^{(1)},\bm\sigma^{(2)}) \rangle_{t,a} + \SOMMA{k=2}{P} \begin{pmatrix}P\\k\end{pmatrix} \langle (q_{12}(\bm\sigma^{(1)},\bm\sigma^{(2)})-\bar{q}_a)^k \rangle_{t,a} \bar{q}_a^{P-k}+(1-P) \bar{q}_a^{P} ,
\end{array}  
\label{eq:RS_pq_Potenziali}
\end{equation}
we obtain, in the thermodynamic limit, under Assumption \ref{eq:seconda},
\begin{equation}\label{eq:zozzona}
 \begin{array}{lll}
     \langle q_{12}(\bm\sigma^{(1)},\bm\sigma^{(2)})^P \rangle_{t,a}  &=&
     P\,\bar{q}_a^{P-1}\langle q_{12}(\bm\sigma^{(1)},\bm\sigma^{(2)}) \rangle_{t,a} -(P-1) \bar{q}_a^{P}
\end{array}  
\end{equation}
for $a=1,2$. Finally, we insert the above in \eqref{eq:compact} and we 
choose the constants $A_1,\ A_2, \ \psi$ in such a way that 
the terms dependent on $\langle q_{12}(\bm\sigma^{(1)},\bm\sigma^{(2)}) \rangle_{t,a}$ cancel out, 
\begin{equation}
\label{eq:values}
    \begin{array}{lll}
         &\psi=\beta '\dfrac{P}{2M}(1+\rho)^{P/2}\n^{P-1} ,
         \\\\
         &A_1^{\2}=\beta '\dfrac{P}{2}\dfrac{(1+\sqrt{\rho_P})^2}{(1+\rho)^{P}}\dfrac{KP!}{2N^{P-1}}\q_1^{^{P-1}},
         \\\\
         &A_2^{\2}=\beta '\dfrac{P}{2}\dfrac{(1+\sqrt{\rho_P})^2}{(1+\rho)^{P}}\dfrac{KP!}{2N^{P-1}}(\q_2^{^{P-1}}-\q_1^{^{P-1}})
    \end{array}
\end{equation}
This makes $d\mathcal{F}^{(P)}/dt$ $t$-independent and 
leads to the thesis \eqref{eq:streaming_RS_Guerra}.

\subsection{Proof of Corollary \ref{cor:nulltemp_rho}}
\label{app:proofnulltemp}

Let us start from the self consistency equations in the $M\to\infty$ limit introduced in Corollary \ref{cor:large},  
we recognize that as $\tilde\beta\to\infty$, we have $\q_2\to 1$, therefore in order to perform the limit we will introduce  the reparametrization
\begin{equation}
\begin{array}{lll}
     \q_2=1-\dfrac{\delta\q_2}{\tilde\beta}&\mathrm{as}&\tilde\beta\to\infty.
\end{array}
\end{equation}
It is now useful to insert an additional term $\tilde\beta y$ in the expression of $g(\beta, \gamma, \bm Y)$ in \eqref{eq:g_of_unsuper_n_app_unsup_troncata}, which now reads as
 \begin{align}
    g(\beta, \gamma, \bm Y)=&\tilde\beta \dfrac{P}{2}\m^{^{P-1}} +\tilde\beta Y^{(2)} \sqrt{
             \gamma\dfrac{P}{2}(1+\sqrt{\rho_P})^2 (\q_2^{^{P-1}} - \q_1 ^{^{P-1}})} \notag \\
             &+ \tilde\beta Y^{(1)}\sqrt{\rho\left(\dfrac{P}{2}\m^{^{P-1}}\right)^2+\gamma
    \dfrac{P}{2}(1+\sqrt{\rho_P})^2\,\q_1^{^{P-1}}\;}\; + \tilde\beta y.
    \label{eq:g_zeroT_rho}
 \end{align}
 \normalsize
Using this new parameter $y$, we can recast the equation for $\q_2$ as a derivative of the magnetization
\begin{equation}
\begin{array}{lll}
     \dfrac{\partial\m}{\partial y}=\delta\q_2-D\Delta\q\Longrightarrow\delta\q_2=\dfrac{\partial\m}{\partial y}+D\Delta\q
\end{array}
\end{equation}
where we have used $\Delta\q=\q_2-\q_1$ and, as $\tilde\beta\to\infty$,  $\tilde\beta\theta\to D \in\mathbb{R}$. Thus, in the zero temperature limit the last three equations in Eq. \eqref{eq:selfLarge} become
\begin{equation}
\begin{array}{lll}
     \bar{m}&\to& \mathbb{E}_{Y^{(1)}} \left\lbrace \dfrac{\mathbb{E}_{Y^{(2)}} \left[\mathrm{sign}{[g(\boldsymbol{Y},\m,\Delta\q)]}\:e^{D|g(\boldsymbol{Y},\m,\Delta\q)|}\,\right]}{\mathbb{E}_{Y^{(2)}} \left[e^{D|g(\boldsymbol{Y},\m,\Delta\q)|}\right]}\right\rbrace \\
\Delta\q &\to& 1-\mathbb{E}_{Y^{(1)}} \left\lbrace \dfrac{\mathbb{E}_{Y^{(2)}} \left[\mathrm{sign}{[g(\boldsymbol{Y},\m,\Delta\q)]}\:e^{D|g(\boldsymbol{Y},\m,\Delta\q)|}\,\right]}{\mathbb{E}_{Y^{(2)}} \left[e^{D|g(\boldsymbol{Y},\m,\Delta\q)|}\right]}\right\rbrace^2 \\
\q_2 &\to& 1 .
\end{array}
\end{equation}
Now, if we suppose $\Delta \q\ll 1$ the \eqref{eq:g_zeroT_rho} reduces to
\begin{equation}
\label{gmodified_rho}
    g(\boldsymbol{Y},\m,\Delta\q)=C_0+C_1 Y^{(1)}+C_2 Y^{(2)}+\mathcal{O}(\Delta\q)
\end{equation}
where
\begin{equation}
    \begin{array}{lll}
         C_0=\dfrac{P}{2}\m^{P-1}\,,
         &
         C_1=\sqrt{\rho\left(\dfrac{P}{2}\m^{P-1}\right)^2+\gamma(1+\sqrt{\rho_P})^2\dfrac{P }{2}}\,,
         &
         C_2=\sqrt{\gamma(1+\sqrt{\rho_P})^2\dfrac{P(P-1)}{2}\Delta\q}\,.
    \end{array}
    \label{eq:A_parameters_RSB_rho}
\end{equation}
Performing the integral over $Y^{(2)}$ we reach Eqs. \eqref{eq:m_beta_rho}-\eqref{eq:q_beta_rho}. 


\end{document}